\documentclass[journal]{IEEEtran}

\usepackage{amsmath,amssymb,mathtools}
\usepackage{graphicx}
\usepackage{xcolor}
\usepackage{tikz}
\usetikzlibrary{arrows.meta,positioning}
\usepackage{cite}
\allowdisplaybreaks

\newcommand{\R}{\mathbb{R}}
\newcommand{\N}{\mathbb{N}}
\newcommand{\E}{\mathbb{E}}
\renewcommand{\Pr}{\mathbb{P}}
\newcommand{\Inf}{\mathsf{Inf}}
\newcommand{\Fin}{\mathsf{Fin}}
\newcommand{\Inv}{\mathsf{Inv}}
\newcommand{\Str}{\mathsf{Str}}
\newcommand{\Par}{\mathsf{Par}}
\newcommand{\AP}{\mathsf{AP}}
\newcommand{\tr}{\operatorname{tr}}
\newcommand{\dist}{\operatorname{dist}}
\newcommand{\dd}{\mathrm{d}}
\newcommand{\one}{\mathbf{1}}
\newcommand{\Snext}{\bigcirc}
\newcommand{\Until}{\mathbin{\mathcal{U}}}
\newcommand{\SZ}{\mathsf{S}_Z}
\newcommand{\SL}{\mathsf{S}_\Lambda}
\newcommand{\Sphi}{\mathsf{S}_\varphi}
\newcommand{\SPsi}{\mathsf{S}_\Psi}
\usetikzlibrary{arrows.meta,shapes.geometric}
\newtheorem{theorem}{Theorem}
\newtheorem{lemma}{Lemma}
\newtheorem{proposition}{Proposition}
\newtheorem{corollary}{Corollary}
\newtheorem{definition}{Definition}
\newtheorem{assumption}{Assumption}
\newtheorem{example}{Example}
\newtheorem{problem}{Problem}
\newtheorem{remark}{Remark}
\makeatletter
\def\section{\@startsection{section}{1}{\z@}{2.0ex plus 1ex minus 1ex}%
{0.5ex plus 0.5ex minus 0ex}{\normalfont\normalsize\centering\scshape}}
\def\subsection{\@startsection{subsection}{2}{\z@}{2.0ex plus 1ex minus 1ex}%
{0.5ex plus 0.5ex minus 0ex}{\normalfont\normalsize\itshape}}
\makeatother

\begin{document}

\title{Complete Certificates for Temporal Specifications\\ on Stochastic Differential Equations}
\author{Arash Bahari Kordabad$^{1}$, Rupak Majumdar$^{1}$, Mohammad H. Mamduhi$^{2}$, and Sadegh Soudjani$^{1,2}$%
\thanks{$^{1}$Max Planck Institute for Software Systems,
Kaiserslautern, Germany. Corresponding author's email:
\texttt{arashbk@mpi-sws.org}}%
\thanks{$^{2}$School of Computer Science, University of Birmingham,
Birmingham, United Kingdom.}%
}

\maketitle
\begin{abstract}
We provide sound and complete certificates for verifying that the trajectories of a stochastic differential equation (SDE) satisfy a temporal requirement with probability one. We first consider recurrence, fairness, and more generally any Boolean combination of requirements that a trajectory visit a closed target set at least once or at arbitrarily large times. Fixed-time sampled trajectories may miss visits between sampling instants. We therefore construct a discrete-time Markov process, the \emph{interval-hit process}, that records which target sets are visited during each sampling interval. We then address general linear temporal logic (LTL) specifications, in which the \emph{next} operator is challenging in continuous time because there is no immediate next time instant. We interpret this operator as referring to the next change in the truth values of atomic propositions. However, labels determined only by the current state may switch infinitely often at a boundary, making this interpretation ill-defined. We therefore introduce a hysteretic label with finitely many switches on bounded intervals and sample the SDE at these switches to obtain an \emph{event-indexed Markov process} that preserves LTL satisfaction exactly under this labelling. In the interval-hit and event-indexed processes, the requirements become Streett acceptance conditions. We reduce these conditions to invariance of a region, safety, and reachability conditions, and certify the first two by supermartingale functions, and reachability by a drift and a variant function. We prove that the resulting certificates are sound and complete. We demonstrate the approach in four examples.
\end{abstract}
\begin{IEEEkeywords}
Certificates,
Continuous-time systems,
Linear temporal logic,
Markov processes,
Qualitative satisfaction,
Stochastic differential equations,
Verification.
\end{IEEEkeywords}

\section{Introduction}
\label{sec:introduction}

\IEEEPARstart{S}{tochastic} differential equations (SDEs) model systems that evolve in continuous time under random disturbances. Linear temporal logic (LTL) expresses requirements on their trajectories, such as visiting a target set at arbitrarily large times or visiting several sets in a prescribed order. We ask how to certify that such a requirement holds with probability one from a given initial state. We seek certificates consisting of functions that satisfy inequalities whose existence is sufficient and, for a given initial state, necessary for the requirement, without reasoning directly about trajectories.

For discrete-time Markov processes, certificate-based methods have been developed for almost-sure termination~\cite{mciver2018new,majumdar2025sound},
quantitative termination~\cite{ChatterjeeEtAl2022},
almost-sure reachability~\cite{majumdar2024necessary,labbadi2026formal},
persistence and recurrence~\cite{chakarov2016deductive}, branching-time temporal specifications~\cite{dimitrova2016probabilistic}, and $\omega$-regular properties~\cite{abate2024stochastic,abate2026complete}. For a prescribed specification and class of processes, a certificate rule is sound if the existence of a certificate implies satisfaction of the specification, and complete if satisfaction implies the existence of a certificate. For linear systems with additive noise, the existence of certificates for almost-sure reachability is characterized by conditions on the system matrices~\cite{kordabad2026certificates}. On countable state spaces, sound and complete rules are known for probabilistic termination~\cite{majumdar2025sound}, and completeness has also been established for Streett certificates~\cite{abate2026complete}. On general state spaces, a rule for almost-sure reachability combines an inequality on the expected value of a drift function with a probabilistic decrease condition on a separate variant function. This rule is sound and, for each initial state, complete~\cite{labbadi2026formal}.
Certificates for computing lower bounds on satisfaction of temporal properties
have also been studied. Relevant approaches include both model-based
methods~\cite{JagtapSoudjaniZamani2021} and data-driven
techniques~\cite{MazouzEtAl2024Permissible,SalamatiEtAl2024,SchonEtAl2026,CasablancaEtAl2026LUCID}.
Extension of these data-driven methods to almost-sure reachability has also been
studied using kernel-based methods~\cite{MajumdarRubabSoudjani2026}.

For continuous-time systems, Lyapunov functions can give necessary and sufficient conditions for a finite expected time for a diffusion to reach a
bounded open set, under nondegeneracy and a regular boundary~\cite[Thm.~3.11]{khasminskii2012stochastic}. Barrier certificates bound
the probability of ever reaching an unsafe set~\cite{prajna2007framework}.
Over a finite horizon, barrier certificates give upper bounds on the
probability of leaving a safe set~\cite{SantoyoDutreixCoogan2021} and
lower bounds on the probability of satisfying safe-LTL specifications
for switched stochastic systems~\cite{anand2019verification}.
Reach-avoid probabilities of controlled diffusions are characterized by
partial differential equations (PDEs)~\cite{mohajerinesfahani2016stochastic},
polynomial barrier conditions certify lower bounds on infinite-horizon
reach-avoid probabilities for SDEs~\cite{Xue2026ReachAvoid}, and reachability
probabilities of linear switching diffusions are computed with certified
precision~\cite{laurenti2017reachability}.
Almost-sure reachability of bounded open sets for SDEs is certified in~\cite{kordabad2026almost}.
Each of these results is restricted to open target sets, to a finite horizon, or to a bound on a probability.
To the best of our knowledge, a certificate that is sound and complete for the almost-sure satisfaction of a general LTL specification by an SDE over an infinite horizon is open. This paper provides such a certificate.

A standard way to define LTL on a continuous-time trajectory, used by~\cite{kloetzer2006fully,wongpiromsarn2016automata,selvaratnam2022sampling}, records its initial label and then each new label whenever the label changes. This word is defined when the label signal is right-continuous and changes only finitely often on every bounded interval. For an SDE, both this word and sampling at fixed times raise difficulties.
First, a trajectory can enter a target set and leave it between two consecutive samples.
For example, a one-dimensional Brownian motion $W$ satisfies $W_t=0$ at arbitrarily large times $t$
almost surely~\cite[Thm.~2.9.6]{karatzas1991brownian}, but, since $\Pr(W_t=0)=0$ for every $t>0$, almost surely none of its
samples at fixed positive times equals $0$. Thus, the sampled process does not
preserve recurrence.
Second, a diffusion can cross a labelling boundary infinitely often immediately
after reaching it, as Brownian motion does at $0$. The successive label changes
then fail to define such a word. Even when the word of label changes is defined, fixed-time sampling can change
whether an LTL formula holds. For example,
$\Box(a\Rightarrow\Snext\neg a)$ means that whenever $a$ is true, it must be false
at the next position in the word. Here, $\Box$ means ``always'' and $\Snext$
means ``next''. In an infinite word of label changes with only one proposition,
true and false alternate, and the formula holds. In a sampled word, two
consecutive samples may both have $a$ true, violating the formula.

Formulas without \emph{next} are invariant under repetitions of a label~\cite[Thm.~7.92]{baier2008principles}, but sampling can also miss label changes, and the word of label changes may not exist.
We address the difficulties described above, for visits to target sets and for labels, by constructing discrete-time Markov processes such that a trajectory satisfies a requirement if and only if the path of the process satisfies the corresponding condition.
For recurrence and fairness, an interval-hit process records which target sets
the trajectory visits between consecutive samples. For every sampling period, a target set is visited at
arbitrarily large times if and only if it is visited in infinitely many sampling
intervals.
For general LTL formulas in which the order of visits matters, and not only which target sets are visited as in recurrence and fairness, for example formulas with \emph{next}, we first construct a hysteretic label. Each proposition becomes true
when the trajectory reaches a closed activation set and false when it reaches
a disjoint closed deactivation set.
Continuity of the trajectory ensures finitely many switches on every bounded
time interval. An event-indexed process then records the state and label at each switch.
The \emph{next} operator therefore refers to the next label change.
If the activation set lies where the proposition is true and the deactivation
set where it is false, the hysteretic label agrees with the memoryless label
on these sets. Outside them, it retains its previous value, and we evaluate LTL
using this label.

On the constructed Markov processes, recurrence and fairness become Streett conditions, while LTL becomes a parity condition that can also be expressed as a Streett condition. The method of~\cite{abate2026complete} reduces Streett verification to invariance, safety and reachability conditions. The invariance certificate bounds the probability that the chain leaves a supporting region. The safety certificates bound the probability of reaching each request set from selected regions by a constant strictly below one. Both use inequalities on the one-step expectations of nonnegative functions. For reachability, we use the rule of~\cite{labbadi2026formal}, which combines a drift function and a variant function. Together, these certificates give sound and, for each initial state, complete verification of the Streett conditions. Each certificate condition involves at most one transition of the constructed Markov chain. Under suitable assumptions on the coefficients and boundaries, the transition probabilities are characterized by parabolic equations for killed diffusions and elliptic equations for exit distributions.

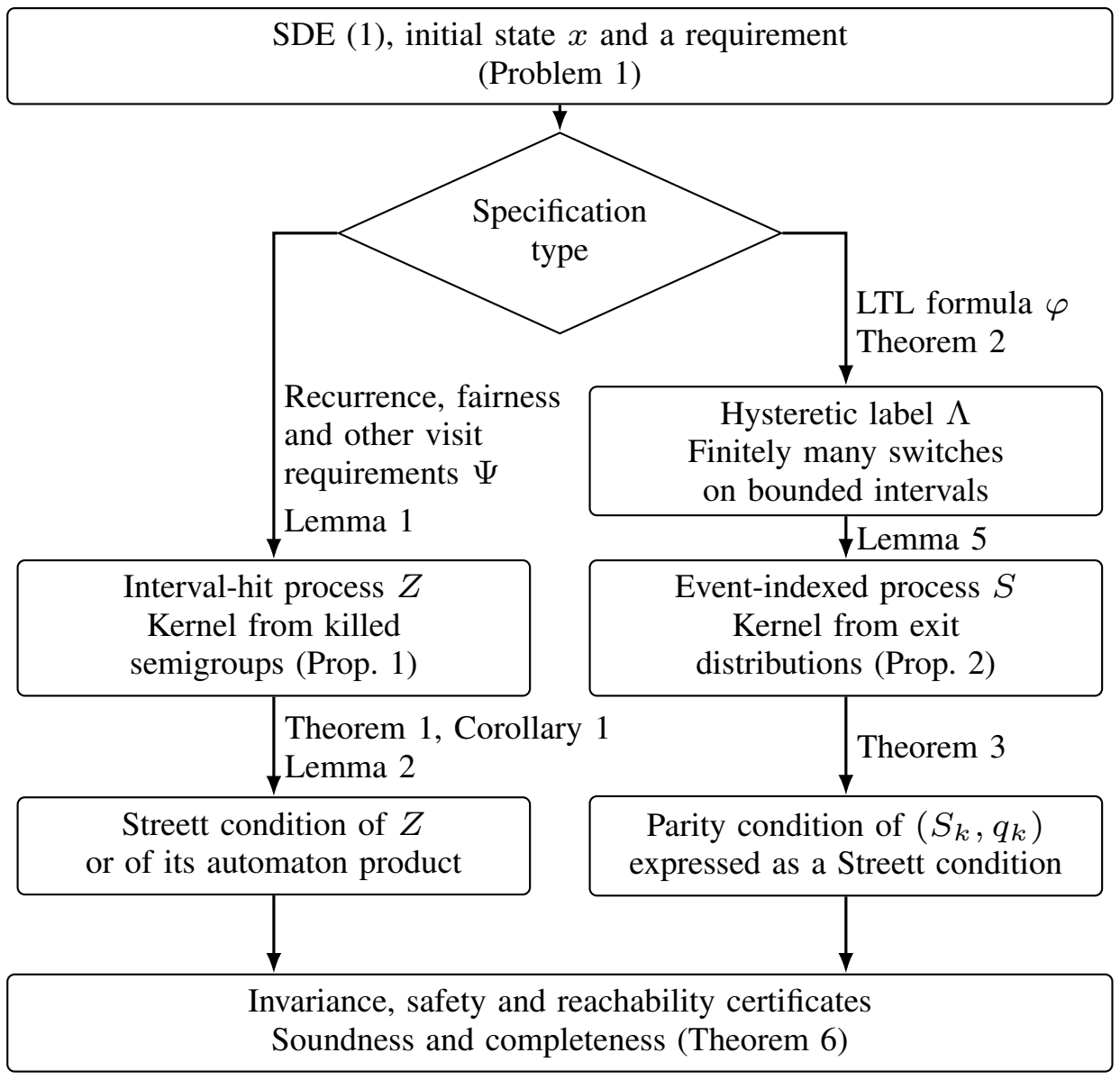
\begin{figure}[t]
  \centering
  \begin{tikzpicture}[
      font=\scriptsize,
      box/.style={
        draw, rounded corners=1.5pt, align=center,
        inner sep=3pt, text width=3.55cm,
        minimum height=0.62cm
      },
      decision/.style={
        draw, diamond, aspect=2.2, align=center,
        inner sep=2pt, text width=1.7cm
      },
      arr/.style={-{Latex[length=1.5mm]}, semithick},
      lab/.style={
        inner sep=2pt, anchor=west,
        font=\scriptsize, align=left
      }
    ]

    \node[box, text width=7.9cm] (sde) at (0,0)
      {SDE~\eqref{eq:sde}, initial state $x$ and a requirement\\
       (Problem~\ref{prob:verification})};

    \node[decision] (type) at (0,-1.3)
      {Specification\\type};

    \node[box] (lam) at (2.1,-2.9)
      {Hysteretic label $\Lambda$\\
       Finitely many switches on bounded intervals};

    \node[box] (z) at (-2.1,-4.2)
      {Interval-hit process $Z$\\
       Kernel from killed semigroups (Prop.~\ref{prop:killed-parabolic})};

    \node[box] (s) at (2.1,-4.2)
      {Event-indexed process $S$\\
       Kernel from exit distributions (Prop.~\ref{prop:exit-elliptic})};

    \node[box] (str) at (-2.1,-5.8)
      {Streett condition of $Z$\\
       or of its automaton product};

    \node[box] (par) at (2.1,-5.8)
      {Parity condition of $(S_k,q_k)$\\
       expressed as a Streett condition};

    \node[box, text width=7.9cm] (cert) at (0,-7.1)
      {Invariance, safety and reachability certificates\\
       Soundness and completeness
       (Theorem~\ref{thm:main})};

    \draw[arr] (sde) -- (type);

    \draw[arr] (type.west) -- (-2.1,-1.3) -- (z.north);
    \node[lab] at (-2.1,-2.95)
      {Recurrence, fairness\\
       and other visit\\ requirements $\Psi$\\[2pt]
       Lemma~\ref{lem:markov-Z}};

    \draw[arr] (type.east) -- (2.1,-1.3) -- (lam.north);
    \node[lab] at (2.1,-1.95)
      {LTL formula $\varphi$\\
       Theorem~\ref{thm:monitor}};

    \draw[arr] (lam) --
      node[lab] {Lemma~\ref{lem:markov-S}} (s);

    \draw[arr] (z) --
      node[lab]
        {Theorem~\ref{thm:recurrence}, Corollary~\ref{cor:emerson-lei}\\
         Lemma~\ref{lem:product-Psi}}
      (str);

    \draw[arr] (s) --
      node[lab] {Theorem~\ref{thm:ltl}} (par);

    \draw[arr] (str.south) -- (cert.north -| str.south);
    \draw[arr] (par.south) -- (cert.north -| par.south);

  \end{tikzpicture}

  \caption{Overview of the approach. The specification type determines
  which Markov process is constructed. Both branches lead to Streett
  conditions, verified through invariance, safety and reachability
  certificates.}
  \label{fig:overview}
\end{figure}

The contributions of this paper are the following.
\begin{itemize}
  \item
  An interval-hit Markov process that, for every sampling period, preserves on every trajectory all finite Boolean combinations of reachability and recurrence events for closed target sets, including every Emerson–Lei condition over these sets.
  \item A hysteretic label with finitely many changes on every bounded time
  interval, and an event-indexed Markov process that preserves every LTL formula
  evaluated on this label, including formulas with the \emph{next} operator.
  \item Sound and, for each initial state, complete Streett certificates for both processes, combining
  invariance and safety certificates with reachability certificates.
  \item Four analytical and numerical examples that illustrate the certificates.
\end{itemize}
Fig.~\ref{fig:overview} summarizes these constructions and their connection
to the certificates presented in this paper.

The rest of the paper is organized as follows. Section~\ref{sec:preliminaries} introduces the model and formulates the problem. Sections~\ref{sec:recurrence} and~\ref{sec:ltl} construct respectively the interval-hit and event-indexed Markov processes, and characterize their transition kernels by parabolic and elliptic equations. Section~\ref{sec:certificates} presents the certificates and proves their soundness and completeness. Section~\ref{sec:examples} illustrates the certificates on four analytical and numerical examples, and Section~\ref{sec:conclusion} concludes the paper. The Appendix contains the proofs.

\section{Preliminaries and Problem Formulation}
\label{sec:preliminaries}
In this section, we introduce the SDE model, the notation for discrete-time Markov processes, and the temporal specifications, and we formulate the problem of characterizing their almost-sure satisfaction.

\subsection{Stochastic System Model}
\label{subsec:model}

We write $\N=\{0,1,2,\dots\}$, $\R_{\ge0}$ and $\R_{>0}$ for the nonnegative and the positive
real numbers, $\one_A$ for the indicator of a set or an event, $A^c$ for the complement of a set $A$,
$\partial A$ and $\dist(x,A)$ for the boundary of $A\subseteq\R^n$ and the Euclidean distance from
$x$ to $A$, and $\lVert\cdot\rVert$ and $\lVert\cdot\rVert_{\mathrm F}$ for the Euclidean and the
Frobenius norms.

Let $(\Omega,\mathcal F,(\mathcal F_t)_{t\ge0},\Pr)$ be a
filtered probability space that satisfies the usual conditions, and let $W_t\in\R^m$ be a standard
Brownian motion with respect to $(\mathcal F_t)$. The system is the time-homogeneous SDE
\begin{equation}
  \dd X_t=f(X_t)\,\dd t+g(X_t)\,\dd W_t,\qquad X_0=x,
  \label{eq:sde}
\end{equation}
where $X_t\in\R^n$ is the state at time $t\ge0$, and $f:\R^n\to\R^n$ and
$g:\R^n\to\R^{n\times m}$ are the coefficients of the SDE.

\begin{assumption}[Locally Lipschitz Coefficients with Linear Growth]
\label{ass:sde}
The coefficients $f$ and $g$ are locally Lipschitz continuous, and there is a constant $c_0>0$
such that for every $x\in\R^n$,
$
  \lVert f(x)\rVert^2+\lVert g(x)\rVert_{\mathrm F}^2\le c_0\bigl(1+\lVert x\rVert^2\bigr).
$
\end{assumption}

Under Assumption~\ref{ass:sde}, the SDE~\eqref{eq:sde} has a pathwise unique strong solution for
every initial state, which does not explode and has continuous
paths~\cite[Thm.~6.22]{klebaner1998introduction},~\cite[Thm.~5.2.5, Rem.~5.3.3]{karatzas1991brownian}.
It is a
strong Markov process~\cite[Prop.~5.3.20, Thm.~5.4.20]{karatzas1991brownian}, also with respect to
$(\mathcal F_t)$, since it is a measurable function of $x$ and
$W$~\cite[Cor.~5.3.23]{karatzas1991brownian} and $W$ starts afresh at every stopping
time~\cite[Thm.~2.6.16]{karatzas1991brownian}.

We write
$\Pr_x$ and $\E_x$ for probability and expectation when $X_0=x$, $(P_tb)(x)=\E_x[b(X_t)]$ for a
bounded or nonnegative Borel function $b$, and
$\mathcal Lb=f^\top\nabla b+\tfrac12\sum_{i,l=1}^n(gg^\top)_{il}\,\partial^2b/\partial x_i\partial x_l$
for the generator acting on twice continuously differentiable $b$. For a closed set
$A\subseteq\R^n$, the first hitting time $\sigma_A=\inf\{t\ge0:X_t\in A\}$, with
$\inf\varnothing=\infty$, is a stopping time, since the paths are continuous and $A$ is closed.
The infimum is attained when it is finite. Thus, a trajectory visits a closed set $A$ at time $t$ when $X_t\in A$.

We write the certificates on discrete-time Markov processes (DTMPs). Let
$\mathsf S$ be a standard Borel space and $K$ a stochastic kernel on it, i.e., for every
$s\in\mathsf S$ the map $A\mapsto K(s,A)$ is a probability measure and for every Borel set $A$
the map $s\mapsto K(s,A)$ is Borel measurable. A process $Y=(Y_k)_{k\in\N}$ with values in
$\mathsf S$ is a DTMP with kernel $K$ if the conditional law of $Y_{k+1}$ given $Y_0,\dots,Y_k$ is
$K(Y_k,\cdot)$ for every $k$. We also call a DTMP a chain. We write $\Pr_s$ and $\E_s$ when
$Y_0=s$, and $\Pr_\mu$ when $Y_0$ has the distribution $\mu$. For every event of the path of $Y$, its
probability under $\Pr_s$ is a Borel function of $s$. This holds for the events
$\{Y_0\in A_0,\dots,Y_k\in A_k\}$, whose probabilities are iterated integrals of $K$, and extends to all events of the
path by the $\pi$--$\lambda$ theorem~\cite[Thm.~2.1.6]{durrett2019probability}. For a Borel function $b$ on $\mathsf S$ with values in $[0,\infty]$, or a bounded one, the one-step expectation is
\begin{equation*}
  (Kb)(s)=\int_{\mathsf S}b(s')\,K(s,\dd s')=\E_s[b(Y_1)].
\end{equation*}
For a Borel set $C\subseteq\mathsf S$ we write $\sigma_C=\inf\{k\ge0:Y_k\in C\}$ and
$\tau_C=\inf\{k\ge1:Y_k\in C\}$, and $\Inf_C$ for the event that $Y_k\in C$ for infinitely many
$k$, with complement $\Fin_C$. With
the SDE and the chains in place, we state the requirements that the certificates must prove.

\subsection{Temporal Specifications}
\label{subsec:specifications}

This paper verifies two families of requirements. The first constrains the times at which the trajectory visits
closed sets, and the second, LTL, constrains the order in which labels occur. For a closed set
$G\subseteq\R^n$, the recurrence event $\Inf(G)$ is the event that $X_t\in G$ for arbitrarily
large $t$, and $\Fin(G)$ is its complement. Since $G$ is closed and the paths are continuous, the
trajectory meets $G$ during $[k,k+1]$ exactly when the infimum of $\dist(X_t,G)$ over the rational
$t$ in $[k,k+1]$ is zero. Hence, $\Inf(G)$ is the event that this happens for infinitely many $k\in\N$, and it is measurable. An Emerson--Lei condition $\Psi$ over closed target sets
$G_1,\dots,G_N$ is a Boolean combination of the events $\Inf(G_j)$ and
$\Fin(G_j)$~\cite[Sec.~3]{emerson1985modalities}, and we write $X\models\Psi$ for the event that it
holds. The usual fairness requirements~\cite{francez1986fairness}, unconditional, weak and strong
fairness, are Emerson--Lei conditions~\cite[Sec.~6]{emerson1985modalities}. A strong-fairness
requirement, also called reactivity or a Streett condition~\cite[Def.~2.3, Rem.~2.4]{abate2026complete},
requires that each response set $F_i$ is visited at arbitrarily large times whenever its request set
$E_i$ is,
\begin{equation}
  \textstyle\bigwedge_{i=1}^{\ell}\bigl(\Inf(E_i)\Rightarrow\Inf(F_i)\bigr),
  \label{eq:streett}
\end{equation}
for closed sets $E_i$ and $F_i$. A generalized B\"uchi condition requires
$\Inf(F_i)$ for every $i$.

The second family is LTL over a finite set $\AP$ of atomic propositions with the alphabet
$\Sigma=2^{\AP}$. A formula $\varphi$ is built from propositions with Boolean connectives and the
temporal operators \emph{next} $\Snext$ and \emph{until} $\Until$, from which \emph{eventually}
$\Diamond$ and \emph{always} $\Box$ follow, and it is interpreted on infinite words
$w=w_0w_1\cdots$ over $\Sigma$ in the standard way~\cite[Sec.~5.1.2]{baier2008principles}. Every LTL formula can be translated into a
nondeterministic B\"uchi automaton~\cite{vardi1986automata},~\cite[Thm.~5.41]{baier2008principles},
and this automaton into a deterministic parity automaton
$\mathcal A_\varphi=(Q,q_{\rm in},\Sigma,\Delta,\kappa)$~\cite{piterman2007nondeterministic}, with a
finite set of states $Q$, an initial state $q_{\rm in}$, a transition function
$\Delta:Q\times\Sigma\to Q$ and priorities $\kappa:Q\to\{0,\dots,\bar\kappa\}$. Its run on $w$ is $q_{\rm in},\Delta(q_{\rm in},w_0),\dots$, and $w\models\varphi$ exactly
when $w$ is accepted, i.e., when the smallest priority that occurs infinitely often in the run is
even.

Interpreting LTL on a trajectory requires a word. A proposition signal $\Lambda$ maps $\R_{\ge0}$
to $\Sigma$. We call it finitely variable if it is right-continuous and has finitely many
discontinuities on every bounded interval. This is the finite variability
of~\cite{selvaratnam2022sampling},~\cite[p.~175]{barringer1986really}, with right-continuity added. For such a signal, the switching times are $\Theta_0=0$ and
$\Theta_{k+1}=\inf\{t>\Theta_k:\Lambda_t\ne\Lambda_{\Theta_k}\}$, and the word
\begin{equation}
  \tr(\Lambda)=\Lambda_{\Theta_0}\Lambda_{\Theta_1}\Lambda_{\Theta_2}\cdots
  \label{eq:trace}
\end{equation}
is called the label-transition trace. The last letter is repeated forever when $\Lambda$ changes only finitely often.
The signal satisfies $\varphi$ when $\tr(\Lambda)\models\varphi$. A memoryless labelling
$L:\R^n\to\Sigma$ produces the signal $t\mapsto L(X_t)$. As we show below, this signal is in general not
finitely variable for a diffusion, and we construct a signal that is.

\subsection{Almost-Sure Verification Problem}
\label{subsec:problem}

Both families lead to the following problem.
\vspace{2mm}

\noindent\fcolorbox{black!35}{black!6}{%
\begin{minipage}{\dimexpr\columnwidth-2\fboxsep-2\fboxrule\relax}
\begin{problem}[Almost-Sure Verification]
\label{prob:verification}
Given the SDE~\eqref{eq:sde}, an initial state $x\in\R^n$ and a temporal specification, we seek a corresponding DTMP and locally checkable conditions on it that ensure that the SDE fulfills the specification. The conditions are on Borel sets and Borel functions on the state space of the DTMP, on functions on $\R_{\ge0}$ and on constants, and each involves at most one transition of the DTMP or only its initial state. A certificate is a tuple of sets, functions and constants that satisfies the conditions, and it must exist if and only if the specification holds almost surely, i.e.,
\begin{enumerate}
  \item[(a)] $\Pr_x(X\models\Psi)=1$, if the specification $\Psi$ is a finite Boolean combination of the events that closed target sets are visited at some time or at arbitrarily large times, and
  \item[(b)] $\Pr_x(\tr(\Lambda)\models\varphi)=1$, if $\varphi$ is an LTL formula, for an
  adapted and finitely variable proposition signal $\Lambda$ generated from $X$.
\end{enumerate}
\end{problem}
\end{minipage}}
\medskip

For a set of initial states, the problem is posed for each of its elements. In part (b), $\Lambda$ is the hysteretic label that we construct for LTL.
Each part is solved in three steps: an exact
discrete-time process, its kernel, and certificates on it. The next section carries out the first two steps for part (a).

\section{Recurrence and Fairness Specifications}
\label{sec:recurrence}

\begin{figure}[t]
  \centering
  \begin{tikzpicture}[x=0.82cm,y=0.45cm,font=\footnotesize]
    \fill[blue!10] (0,1.45) rectangle (6,2.05);
    \draw[blue!55] (0,1.45) -- (6,1.45);
    \draw[blue!55] (0,2.05) -- (6,2.05);
    \node[blue!70!black,anchor=west] at (6.08,1.75) {$G$};
    \foreach \k in {0,...,6}{\draw[dashed,gray!60] (\k,0.12) -- (\k,2.75);}
    \draw[->] (-0.25,0.12) -- (-0.25,2.95) node[above] {$X_t$};
    \draw[thick,smooth] plot coordinates {
      (0,0.42) (0.38,0.82) (0.62,1.74) (0.82,0.88) (1,0.48) (1.5,0.62) (2,0.42)
      (2.32,0.95) (2.55,1.78) (2.78,0.93) (3,0.48) (3.28,1.08) (3.52,1.82) (3.78,0.96)
      (4,0.42) (4.5,0.62) (5,0.48) (5.30,1.05) (5.55,1.76) (5.78,0.91) (6,0.42)};
    \foreach \k/\v in {0/0.42,1/0.48,2/0.42,3/0.48,4/0.42,5/0.48,6/0.42}{\fill (\k,\v) circle (1.5pt);}
    \node[anchor=east] at (-0.18,-0.18) {$t$};
    \node at (0,-0.18) {$0$};
    \node at (1,-0.18) {$\delta$};
    \foreach \k in {2,...,6}{\node at (\k,-0.18) {$\k\delta$};}
    \node[anchor=east] at (-0.18,-1.05) {$\chi_k$};
    \foreach \k/\v in {1/1,2/0,3/1,4/1,5/0,6/1}{
      \draw[gray!65] ({\k-1+0.10},-0.65) -- ({\k-0.10},-0.65);
      \node at ({\k-0.5},-1.05) {$\v$};}
  \end{tikzpicture}
  \caption{Point samples (dots) of a trajectory may miss visits to the target set $G$ (blue strip). The hit vector $\chi_k$, here one bit written under the interval $[(k-1)\delta,k\delta]$, records whether the trajectory visited $G$ during that interval.}
  \label{fig:interval-hit}
\end{figure}
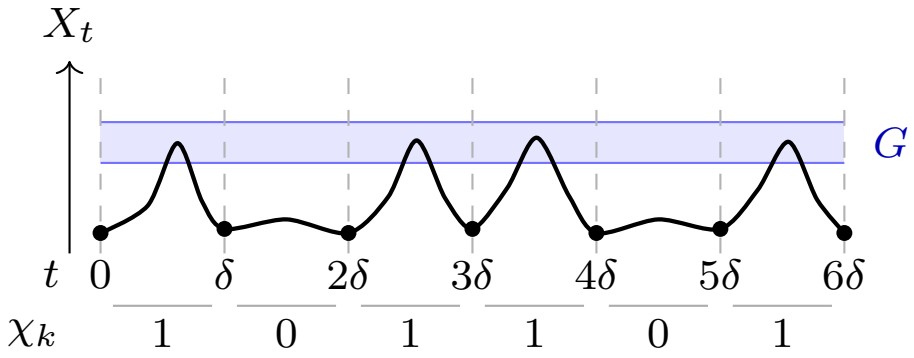

Since fixed-time samples can miss visits, we pair each sample with a record of the target sets that the trajectory visited during the last sampling interval (Fig.~\ref{fig:interval-hit}). This section defines the resulting process, proves that it preserves every visit requirement exactly, and computes its kernel.

\subsection{Interval-Hit Markov Process}
\label{subsec:interval-hit}

Fairness conditions involve several target sets, and the process records the visits to all
of them simultaneously. Let $G_1,\dots,G_N$ be closed target sets and let $\delta>0$ be the sampling period.
The hit vector, as we call it, $\chi_k=(\chi_{k,1},\dots,\chi_{k,N})\in\{0,1\}^N$ records the visits
during the last interval,
\begin{equation}
  \chi_{k+1,j}=\one\bigl\{X_t\in G_j\text{ for some }t\in[k\delta,(k+1)\delta]\bigr\},
  \label{eq:hit-vector}
\end{equation}
for $k\in\N$ and $j=1,\dots,N$, and $\chi_0\in\{0,1\}^N$ is arbitrary. The interval-hit process
is $Z_k=(X_{k\delta},\chi_k)$, with values in $\SZ=\R^n\times\{0,1\}^N$. We call the elements
$\eta$ of $\{0,1\}^N$ hit patterns. For a pattern $\eta$ and a Borel set $A\subseteq\R^n$, the
probability of ending in $A$ with the visits $\eta$ is
$  K^\eta(x,A)=\Pr_x\bigl(X_\delta\in A,\ \chi_1=\eta\bigr),
$
and the kernel of $Z$ acts on a nonnegative or bounded Borel function $b$ on $\SZ$ by
\begin{equation}
  (K_Zb)(x,\eta)=\sum_{\eta'\in\{0,1\}^N}\int_{\R^n}b(y,\eta')\,K^{\eta'}(x,\dd y).
  \label{eq:kernel-Z}
\end{equation}
Thus, $K_Z((x,\eta),A\times\{\eta'\})=K^{\eta'}(x,A)$, i.e., the next sample and the next hit vector are drawn jointly, and neither depends on the previous hit vector $\eta$.

\begin{lemma}[Markov Property of the Interval-Hit Process]
\label{lem:markov-Z}
For every $x\in\R^n$ and $\chi_0\in\{0,1\}^N$, the process $Z$ with $Z_0=(x,\chi_0)$ is a DTMP
with kernel $K_Z$ under $\Pr_x$.
\end{lemma}

We write $\Pr_{(x,\eta)}$ for the law of $Z$ started at $(x,\eta)$. We now relate its hit vectors to recurrence.

\subsection{Exact Preservation of Recurrence}
\label{subsec:recurrence}

For a target set $G_j$, the set $\widetilde G_j=\{(x,\eta)\in\SZ:\eta_j=1\}$ collects the states
whose last interval contains a visit to $G_j$.

\begin{theorem}[Exact Preservation of Recurrence]
\label{thm:recurrence}
For every continuous trajectory, every target set $G_j$ and every sampling period $\delta>0$, the
trajectory visits $G_j$ at arbitrarily large times if and only if $Z_k\in\widetilde G_j$ for
infinitely many $k$. Consequently, for every $x\in\R^n$ and $\chi_0\in\{0,1\}^N$
\begin{equation}
  \Pr_x\bigl(\Inf(G_j)\bigr)=\Pr_{(x,\chi_0)}\bigl(\Inf_{\widetilde G_j}\bigr).
  \label{eq:recurrence-probability}
\end{equation}
\end{theorem}

Thus, recurrence of a target set, including one of Lebesgue measure zero, is recurrence of a set of the interval-hit process.

\subsection{Extension to Visit Requirements and Fairness}
\label{subsec:emerson-lei}

We call a finite Boolean combination $\Psi$ of the events $\{\sigma_{G_j}<\infty\}$ and $\Inf(G_j)$ a visit requirement. Every Emerson--Lei condition is one, and so are reachability $\{\sigma_{G_j}<\infty\}$ and safety $\{\sigma_{G_j}=\infty\}$. Let $\widetilde\Psi$ denote the condition on $Z$ obtained by replacing every $\{\sigma_{G_j}<\infty\}$ by $\{\sigma_{\widetilde G_j}<\infty\}$, every $\Inf(G_j)$ by $\Inf_{\widetilde G_j}$ and every $\Fin(G_j)$ by $\Fin_{\widetilde G_j}$.

\begin{corollary}[Exact Preservation of Visit Requirements]
\label{cor:emerson-lei}
For the zero hit pattern $\chi_0=\mathbf 0$ and every continuous trajectory, $X\models\Psi$ if and only if $Z\models\widetilde\Psi$. Consequently, $\Pr_x(X\models\Psi)=\Pr_{(x,\mathbf 0)}(Z\models\widetilde\Psi)$ for every $x$ and every $\delta>0$. For an Emerson--Lei condition, both statements hold for every $\chi_0$.
\end{corollary}

Many fairness requirements are already Streett conditions on $Z$, without an automaton. A sequence
visits a union of sets infinitely often exactly when it visits one of the sets infinitely often. A
clause $\Fin(G_{j_0})\vee\bigvee_{j:\zeta_j=1}\Inf(G_j)$, with a target index $j_0$ and a pattern
$\zeta\in\{0,1\}^N$, therefore becomes, on $Z$, the event
$\Fin_{\widetilde G_{j_0}}\cup\Inf_{\bigcup_{j:\zeta_j=1}\widetilde G_j}$, i.e., the Streett pair
$(\widetilde G_{j_0},\bigcup_{j:\zeta_j=1}\widetilde G_j)$, and a clause without a $\Fin$ literal
gives a pair whose first set is $\SZ$. Hence, every Emerson--Lei condition with a conjunctive normal
form that has at most one $\Fin$ literal per clause is, on $Z$, a Streett condition of the
form~\eqref{eq:streett}, with one pair per clause and with events of $Z$ in place of events of $X$.
With the sets $E_i$ and $F_i$ among the target sets $G_1,\dots,G_N$, strong
fairness~\eqref{eq:streett} has the pairs $(\widetilde E_i,\widetilde F_i)$, and a generalized
B\"uchi condition the pairs $(\SZ,\widetilde F_i)$. For every other visit requirement, $\widetilde\Psi$ is an LTL formula over the hit vectors, a Boolean combination of \emph{eventually} $\eta_j=1$ and \emph{always eventually} $\eta_j=1$, which are $\{\sigma_{\widetilde G_j}<\infty\}$ and $\Inf_{\widetilde G_j}$. This formula is therefore recognized by a
deterministic parity automaton, as every LTL formula is (Section~\ref{subsec:specifications}), and a
parity condition is a Streett condition, as Section~\ref{sec:certificates} shows. We write
$\mathcal A_\Psi=(Q_\Psi,q^\Psi_{\rm in},\{0,1\}^N,\Delta_\Psi,(E^\Psi_i,F^\Psi_i)_i)$ for this
automaton with its Streett pairs. It accepts exactly the words $\chi_0\chi_1\cdots$ on which
$\widetilde\Psi$ holds.

\begin{lemma}[Automaton Product]
\label{lem:product-Psi}
The process $(Z_k,q_k)$ with $q_0=\Delta_\Psi(q^\Psi_{\rm in},\chi_0)$ and
$q_{k+1}=\Delta_\Psi(q_k,\chi_{k+1})$ is a DTMP on $\SPsi=\SZ\times Q_\Psi$, and its Streett
condition with the pairs $(\SZ\times E^\Psi_i,\SZ\times F^\Psi_i)$ holds on every trajectory
exactly when $Z\models\widetilde\Psi$.
\end{lemma}

The kernel of the product is
$(K_\Psi b)((x,\eta),q)=\sum_{\eta'}\int_{\R^n}b((y,\eta'),\Delta_\Psi(q,\eta'))\,K^{\eta'}(x,\dd y)$. It remains to compute the kernel $K_Z$, on
which the certificate conditions depend.

\subsection{Kernel Characterization}
\label{subsec:kernel-Z}
The kernel of the interval-hit process consists of the probabilities that a trajectory ends in a given set and visits exactly the specified target sets during one sampling interval, which we compute from probabilities of avoiding selected target sets.

For a pattern $\zeta\in\{0,1\}^N$, let $G_\zeta$ be the union of
the target sets $G_j$ with $\zeta_j=1$, and define
\begin{equation*}
  (P^\zeta_t b)(x)
  =\E_x\bigl[b(X_t)\,\one_{\{\sigma_{G_\zeta}>t\}}\bigr].
\end{equation*}
This is the semigroup killed on entering $G_\zeta$, and $(P^\zeta_t\one_A)(x)$ is the probability that a trajectory from $x$ ends in $A$ at time $t$ without visiting $G_\zeta$.
For the zero pattern, $G_{\mathbf 0}=\varnothing$ and
$P^{\mathbf 0}_t=P_t$.  For two patterns we
write $\zeta\le\eta$ componentwise, $\lvert\zeta\rvert$ for the number of ones, and $\eta^c$ for
the pattern with $\eta^c_j=1-\eta_j$. For $\zeta\le\eta$, the componentwise sum $\eta^c+\zeta$ is
again a pattern, since $\zeta_j=1$ implies $\eta^c_j=0$.

\begin{lemma}[Inclusion--Exclusion Kernel]
\label{lem:inclusion-exclusion}
For every hit pattern $\eta$, every $x\in\R^n$ and every Borel set $A$,
\begin{equation}
  K^\eta(x,A)=\sum_{\zeta\le\eta}(-1)^{\lvert\zeta\rvert}\bigl(P^{\eta^c+\zeta}_\delta\one_A\bigr)(x).
  \label{eq:inclusion-exclusion}
\end{equation}
\end{lemma}

To evaluate the kernel in Lemma~\ref{lem:inclusion-exclusion},
we characterize the killed semigroups through parabolic PDEs.

\begin{proposition}
\label{prop:killed-parabolic}
Let Assumption~\ref{ass:sde} hold and let the diffusion be uniformly elliptic, i.e., every eigenvalue
of $g(x)g(x)^\top$ is at least a constant $\lambda>0$ for every $x\in\R^n$. Then, for every pattern $\zeta$:
\begin{enumerate}
  \item[(a)] For every $x\in\R^n$ and $t>0$, the measures $A\mapsto\Pr_x(X_t\in A)$ and
  $A\mapsto\Pr_x(X_t\in A,\ \sigma_{G_\zeta}>t)$ have densities $p_t(x,\cdot)$ and
  $p^\zeta_t(x,\cdot)$, respectively, with $0\le p^\zeta_t(x,\cdot)\le p_t(x,\cdot)$ almost
  everywhere, and
  $
    (P^\zeta_t b)(x)=\int_{\R^n}b(y)\,p^\zeta_t(x,y)\,\dd y
$  for every nonnegative Borel function $b$.

  \item[(b)] Let $D$ be a bounded open set with a $C^2$ boundary, and let $b$ be continuous on
  $\overline D$ and zero on $\partial D$. Then $u(t,x)=\E_x[b(X_t)\one_{\{\sigma_{D^c}>t\}}]$ is the
  unique function that is continuously differentiable once in $t$ and twice in $x$ on
  $(0,\delta]\times D$, continuous on $[0,\delta]\times\overline D$, and satisfies
  \begin{equation*}
    \begin{cases}
      \partial_tu=\mathcal Lu,
        &(t,x)\in(0,\delta]\times D,\\
      u(t,x)=0,
        &(t,x)\in(0,\delta]\times\partial D,\\
      u(0,x)=b(x),
        &x\in D.
    \end{cases}
  \end{equation*}
\end{enumerate}
\end{proposition}

The Appendix proves Proposition~\ref{prop:killed-parabolic} with the transition density of~\cite[Ch.~6, Thm.~5.4]{friedman1975stochastic} for (a) and the stochastic representation in~\cite[Ch.~6, Thm.~5.2]{friedman1975stochastic} for (b).

By (a) and Lemma~\ref{lem:inclusion-exclusion}, $K^\eta(x,\cdot)$ has the density
$q^\eta(x,\cdot)=\sum_{\zeta\le\eta}(-1)^{\lvert\zeta\rvert}p^{\eta^c+\zeta}_\delta(x,\cdot)$ almost
everywhere, and $(K_Zb)(x,\eta)=\sum_{\eta'}\int_{\R^n}b(y,\eta')\,q^{\eta'}(x,y)\,\dd y$ for every
nonnegative Borel function $b$. For $D=G_\zeta^c$, part (b) gives $u=P^\zeta_tb$, for $b$ as in (b), when $G_\zeta^c$ is
bounded with a $C^2$ boundary. Without this restriction, for continuous $b\ge0$,
$P^\zeta_tb$ is the increasing limit of the functions $u$ of (b) on bounded smooth domains $D_m$ that
exhaust $G_\zeta^c$, with data that vanish on $\partial D_m$ and increase to $b$, by monotone
convergence, since a trajectory that avoids $G_\zeta$ on $[0,t]$ stays in a compact subset of
$G_\zeta^c$. The next
section treats LTL.

\section{Linear Temporal Logic Specifications}
\label{sec:ltl}

LTL is evaluated on the sequence of labels generated by a trajectory. For a diffusion, however, memoryless labelling may lead to infinitely many label changes in a finite time interval, making the corresponding word undefined. We address this problem using hysteretic labelling, which produces a well-defined word, and construct a DTMP by sampling the SDE at label-switching times. We prove that this process preserves LTL satisfaction exactly and characterize its transition kernel.

Memoryless labelling assigns a letter to each state through
a Borel map $L:\R^n\to\Sigma$.
The truth set of proposition $a$ is
$A_a=\{x:a\in L(x)\}$, with boundary $\partial A_a$.

\begin{figure}[t]
  \centering
  \includegraphics[scale=1]{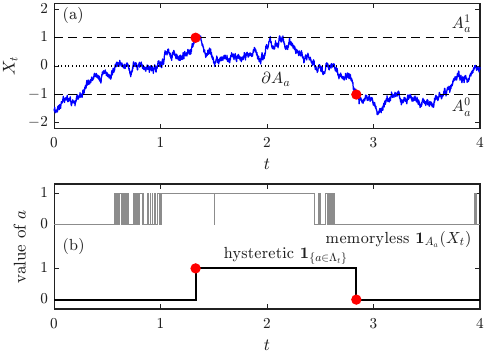}
  \caption{Memoryless and hysteretic labelling of one simulated path of the Brownian motion of
  Example~\ref{ex:chattering}. (a) The path (blue), the boundary $\partial A_a$ of the truth set $A_a=[0,\infty)$
  (black, dotted) and the boundaries of the switching sets $A^1_a=[1,\infty)$ and
  $A^0_a=(-\infty,-1]$ (black, dashed). (b) The memoryless label $\one_{A_a}(X_t)$ (gray), which chatters after the first visit to $0$, and the hysteretic label (black). Red dots mark the hysteretic switches in both panels.}
  \label{fig:hysteresis}
\end{figure}

\begin{example}[Boundary Chattering]
\label{ex:chattering}
Consider the one-dimensional SDE $\dd X_t=\dd W_t$ with $X_0=-1.5$ and $A_a=[0,\infty)$. By the strong Markov property, conditional on $\sigma_{\{0\}}<\infty$, the process $X_{\sigma_{\{0\}}+t}-X_{\sigma_{\{0\}}}$ is a Brownian motion started at zero. Such a process takes both positive and negative values in every interval $(0,t)$ with $t>0$, almost surely, by the law of the iterated logarithm~\cite[Thm.~2.9.23]{karatzas1991brownian}. Hence, the switching times of $\one_{A_a}(X_t)$ accumulate at $\sigma_{\{0\}}$ from the right, and the label-transition trace~\eqref{eq:trace} is undefined (Fig.~\ref{fig:hysteresis}).
\end{example}

\subsection{Hysteretic Labelling and Finite Variability}
\label{subsec:monitor}

For each proposition $a$, the designer chooses two closed sets, the activation set $A^1_a$ and the
deactivation set $A^0_a$.

\begin{assumption}[Closed Disjoint Switching Sets]
\label{ass:switching}
For every $a\in\AP$, the sets $A^1_a$ and $A^0_a$ are closed and disjoint.
\end{assumption}

The hysteretic label stores its current letter $h\in\Sigma$.
A letter $h$ is consistent with a state $x$ if every proposition
whose activation set contains $x$ is on, and every proposition
whose deactivation set contains $x$ is off. We denote the set
of such letters by $\Sigma(x)$.  Since, by
Assumption~\ref{ass:switching}, $A^1_a$ and $A^0_a$ are disjoint, each proposition can be assigned a value consistent with $x$: it must be on in $A^1_a$, off in $A^0_a$, and can take either value elsewhere. Hence, $\Sigma(x)$ is nonempty for every $x$. Starting from a letter $h$, a proposition switches on when
the trajectory reaches its activation set $A^1_a$, and switches
off when it reaches its deactivation set $A^0_a$. Otherwise,
its value remains unchanged. We return to Example~\ref{ex:chattering} with the switching sets
$A^1_a=[1,\infty)$ and $A^0_a=(-\infty,-1]$. In Fig.~\ref{fig:hysteresis}(b), the hysteretic label
of the simulated path (black) switches only twice, whereas its memoryless label (gray) changes many
times. Define the set $B_h$ that collects all states
that trigger a switch, and $\Gamma(h,y)$ that gives the updated
letter at state $y$:
\begin{align*}
  B_h&=\textstyle\bigcup_{a\notin h}A^1_a\ \cup\ \bigcup_{a\in h}A^0_a,\\
  \Gamma(h,y)&=\{a\in h:y\notin A^0_a\}\cup\{a\notin h:y\in A^1_a\}.
\end{align*}
The set $B_h$ is closed, $h$ is consistent with $y$ if
and only if $y\notin B_h$, and $\Gamma(h,y)$ is
consistent with $y$.

For an initial state $x$ and a consistent letter
$h\in\Sigma(x)$, set $\Theta_0=0$ and $\Lambda_0=h$.
We define the hysteretic label recursively by
\begin{align}
  \Theta_{k+1}&=\inf\{t>\Theta_k:X_t\in B_{\Lambda_{\Theta_k}}\},\nonumber\\*
  \Lambda_t&=\Lambda_{\Theta_k}\quad\text{for }t\in[\Theta_k,\Theta_{k+1}),\label{eq:monitor}\\*
  \Lambda_{\Theta_{k+1}}&=\Gamma\bigl(\Lambda_{\Theta_k},X_{\Theta_{k+1}}\bigr).\nonumber
\end{align}
The next theorem
shows that the finite switching times are strictly increasing and do not accumulate, so these rules define $\Lambda_t$ for all $t\ge0$. The times $\Theta_k$ are the switching times in~\eqref{eq:trace}, since $\Gamma(h,y)\ne h$ for every $y\in B_h$.

We write $\Pr_{(x,h)}$ for the law of $(X,\Lambda)$ with
$X_0=x$ and $\Lambda_0=h$. The initial letter
can also be fixed by $h=\{a\in\AP:x\in A^1_a\}$, which belongs
to $\Sigma(x)$ by Assumption~\ref{ass:switching}. With this
choice, the verification problem depends only on $x$.

\begin{theorem}[Well-Posed Hysteretic Label]
\label{thm:monitor}
Let Assumption~\ref{ass:switching} hold and let $h\in\Sigma(x)$. For every continuous trajectory
started at $x$:
\begin{enumerate}
  \item[(a)] $\Lambda_t\in\Sigma(X_t)$ for every $t\ge0$.
  \item[(b)] $\Theta_{k+1}>\Theta_k$ whenever $\Theta_k<\infty$, every bounded interval contains
  finitely many switching times, and $\Lambda$ is finitely variable.
  \item[(c)] If $A^1_a\subseteq A_a$ and $A^0_a\subseteq A_a^c$, then $a\in\Lambda_t$ if and only
  if $a\in L(X_t)$, at every $t$ with $X_t\in A^1_a\cup A^0_a$.
\end{enumerate}
Moreover, every $\Theta_k$ is a stopping time, and $\Lambda$ is adapted.
\end{theorem}

We construct the switching sets from the truth sets of the memoryless label by introducing a band width.

\begin{lemma}[Switching Sets from a Band Width]
\label{lem:band}
For a set $A_a\subseteq\R^n$ and $\rho_a>0$, the sets $A^1_a=\{x:\dist(x,A_a^c)\ge\rho_a\}$ and
$A^0_a=\{x:\dist(x,A_a)\ge\rho_a\}$ are closed, lie in $A_a$ and in $A_a^c$, respectively, are at
distance at least $\rho_a$ from each other, and every $x$ with $\dist(x,\partial A_a)\ge\rho_a$ lies
in one of them.
\end{lemma}

Theorem~\ref{thm:monitor}(c) states that the hysteretic label agrees with the
memoryless one at distance $\rho_a$ or more from the boundary of $A_a$. The band of points at distance less than $\rho_a$ from the boundary is part of the specification, like the margin of robust
semantics~\cite{fainekos2009robustness}. The hysteretic label has a well-defined word, and the next subsection constructs a DTMP that
generates it.

\subsection{Event-Indexed Markov Process}
\label{subsec:event-indexed}

We sample the SDE and the label at the switching times. When no further switch occurs, the process
moves to a terminal state $\partial_h$, whose letter is the current letter $h$. The state space is
\begin{equation*}
  \SL=\{(x,h)\in\R^n\times\Sigma:h\in\Sigma(x)\}\cup\{\partial_h:h\in\Sigma\},
\end{equation*}
and from $(x,h)$ the next state is the point at which the trajectory first reaches $B_h$, together
with the updated letter, or $\partial_h$ when the trajectory never reaches $B_h$, i.e.,
\begin{equation}
  \begin{aligned}
  (K_\Lambda b)(x,h)={}&\E_x\Bigl[b\bigl(X_{\sigma_{B_h}},\Gamma(h,X_{\sigma_{B_h}})\bigr)\one_{\{\sigma_{B_h}<\infty\}}\Bigr]\\*
  &+b(\partial_h)\Pr_x(\sigma_{B_h}=\infty),
  \end{aligned}
  \label{eq:kernel-Lambda}
\end{equation}
while the states $\partial_h$ are terminal, i.e., $K_\Lambda(\partial_h,\{\partial_h\})=1$ for every $h\in\Sigma$. Since $x\notin B_h$, the hitting time $\sigma_{B_h}$ is positive.

\begin{lemma}[Markov Property of the Event-Indexed Process]
\label{lem:markov-S}
Under $\Pr_{(x,h)}$ with $h\in\Sigma(x)$, the process $S_k=(X_{\Theta_k},\Lambda_{\Theta_k})$, with
$S_{k+1}=\partial_{\Lambda_{\Theta_k}}$ and constant afterwards when $\Theta_{k+1}=\infty$, is a DTMP
on $\SL$ with kernel $K_\Lambda$.
\end{lemma}

The process $S$ discards the durations between switches and retains the order of the labels,
which is all that LTL evaluates.

\subsection{Exact Preservation of LTL Satisfaction}
\label{subsec:ltl-exact}

We write $h_k$ for the letter of $S_k$. The following theorem relates LTL satisfaction to parity acceptance.

\begin{theorem}[Exact Preservation of LTL Satisfaction]
\label{thm:ltl}
Let Assumption~\ref{ass:switching} hold and let $h\in\Sigma(x)$. For every continuous trajectory,
$\tr(\Lambda)=h_0h_1h_2\cdots$. Consequently, for every LTL formula $\varphi$,
\begin{equation}
  \Pr_{(x,h)}\bigl(\tr(\Lambda)\models\varphi\bigr)=\Pr_{((x,h),q_0)}(\Par),
  \label{eq:ltl-probability}
\end{equation}
where the right-hand side refers to the DTMP $(S_k,q_k)$ on $\Sphi=\SL\times Q$ with
$q_0=\Delta(q_{\rm in},h_0)$ and $q_{k+1}=\Delta(q_k,h_{k+1})$, and $\Par$ is the event that the
smallest priority $\kappa(q_k)$ that occurs infinitely often is even.
\end{theorem}

\begin{remark}
\label{rem:next}
\upshape
The label-transition trace is stutter-free~\cite{peled1997stutter}, i.e., consecutive letters differ unless switching has stopped, in which case the last letter is repeated forever. Indeed, $\Gamma(h,y)\ne h$ for $y\in B_h$. LTL is evaluated on this word with its standard semantics. Let a letter $h$ denote the formula that holds exactly when the current letter is $h$, and let $\Theta_k<\infty$. For $h'\ne h$, the formula $h\wedge\Snext h'$ holds at position $k$ if and only if the current letter is $h$, a next switch occurs, and it gives the letter $h'$. The formula $h\wedge\Snext h$ holds at position $k$ if and only if the label retains $h$ at every $t\ge\Theta_k$. Thus, \emph{next} refers to the letter after the next switch, no matter at which time $\Theta_{k+1}$ of the trajectory this switch occurs, or to the current letter if no switch follows. For example, with $\AP=\{a,b\}$, at a position labelled $\{a\}$, the formula $a\Until b$ holds if and only if $\Snext b$ holds.
\end{remark}

\subsection{Kernel Characterization}
\label{subsec:kernel-Lambda}

To evaluate the kernel in Lemma~\ref{lem:markov-S}, we need the location of the next switch
and the probability that no further switch occurs. The next switch is the first exit from
$B_h^c$, where the label retains the letter $h$. This exit problem leads to elliptic PDEs,
whereas the fixed-time transition of the interval-hit process leads to the parabolic PDEs
of Proposition~\ref{prop:killed-parabolic}.

\begin{proposition}
\label{prop:exit-elliptic}
Let Assumptions~\ref{ass:sde} and~\ref{ass:switching} hold, fix $h\in\Sigma$, and write $D=B_h^c$.
For a nonnegative or bounded Borel function $b$ on $\SL$, let
$\hat b(y)=b(y,\Gamma(h,y))$ on $\partial D$. Then:
\begin{enumerate}
  \item[(a)] For every $x\in D$,
  \begin{align*}
    u(x)&=\E_x\bigl[\hat b(X_{\sigma_{B_h}})
                       \one_{\{\sigma_{B_h}<\infty\}}\bigr],\\
    (K_\Lambda b)(x,h)&=u(x)+b(\partial_h)\Pr_x(\sigma_{B_h}=\infty).
  \end{align*}
  The same exit expectation with $\hat b=1$ gives
  $\Pr_x(\sigma_{B_h}<\infty)$.

  \item[(b)] Suppose that the diffusion is uniformly elliptic, i.e., $g(x)g(x)^\top\ge\lambda I$ for
  some $\lambda>0$ and every $x\in\R^n$, that $\hat b$ is bounded and
  continuous on $\partial D$, and that $\Pr_x(\sigma_{B_h}<\infty)=1$ for every $x\in D$.
  Assume also that every $y\in\partial D$ is regular~\cite[Def.~9.2.8]{oksendal2003stochastic}, i.e.,
  \begin{equation*}
    \Pr_y\bigl(\inf\{t>0:X_t\in B_h\}=0\bigr)=1.
  \end{equation*}
  Then $u$ is the unique bounded function in $C^2(D)\cap C(\overline D)$ satisfying
  \begin{equation*}
    \begin{cases}
      \mathcal Lu=0,&x\in D,\\
      u(x)=\hat b(x),&x\in\partial D,
    \end{cases}
  \end{equation*}
  and $(K_\Lambda b)(x,h)=u(x)$. Almost-sure exit and the regularity of every $y\in\partial D$ hold
  whenever $D$ is bounded with a $C^2$ boundary.
\end{enumerate}
\end{proposition}

The proof is in the Appendix. Intuitively, the label keeps the letter $h$ until the trajectory first reaches $B_h$, so the next state is determined by the exit point from $D=B_h^c$. The expected value of data evaluated at the exit point satisfies $\mathcal Lu=0$ inside $D$ and equals the data on $\partial D$; for a Brownian motion on an interval, for instance, the probability of leaving through the right end is linear in the initial point.
The automaton coordinate requires no additional computation, since the kernel $K_\varphi$ of the product
$(S_k,q_k)$ satisfies $(K_\varphi b)(s,q)=(K_\Lambda b_q)(s)$ for $s\in\SL$, with
$b_q(y,h')=b((y,h'),\Delta(q,h'))$ and $b_q(\partial_{h'})=b(\partial_{h'},\Delta(q,h'))$.

In the next section, we present certificates for this parity condition, and for the Streett conditions of the previous section.

\section{Complete Certificates for Temporal Specifications}
\label{sec:certificates}

The previous sections represented recurrence, fairness and other visit requirements as Streett acceptance conditions (Theorem~\ref{thm:recurrence}, Corollary~\ref{cor:emerson-lei} and Lemma~\ref{lem:product-Psi}), and LTL satisfaction as parity acceptance (Theorem~\ref{thm:ltl}), on the corresponding DTMPs.

We express parity acceptance as an equivalent Streett condition. Each state $(s,q)\in\Sphi$ of the product DTMP $(S_k,q_k)$ has priority $\kappa(q)$. For each odd priority $2i-1$, define
\begin{equation*}
  \begin{aligned}
    E_i&=\{(s,q)\in\Sphi:\kappa(q)=2i-1\},\\
    F_i&=\{(s,q)\in\Sphi:\kappa(q)<2i-1,\ \kappa(q)\text{ even}\},
  \end{aligned}
\end{equation*}
for $i=1,\dots,\lceil\bar\kappa/2\rceil$, where $\bar\kappa$ is the largest priority. These pairs require that whenever an odd priority occurs infinitely often, a smaller even priority also occurs infinitely often. This holds exactly when the smallest priority occurring infinitely often is even, establishing the equivalence.

It therefore suffices to certify Streett conditions. A Streett certificate combines reachability certificates with invariance and safety certificates~\cite{abate2026complete}. More specifically, the construction of~\cite[Thm.~3.7]{abate2026complete} uses a supporting region $I$ and, for each Streett pair $(E_i,F_i)$, a set $J_i\subseteq I\setminus E_i$. An invariance certificate bounds the probability of leaving $I$. A safety certificate bounds the probability of reaching $E_i$ from any state in $J_i$ by a constant strictly below one. A reachability certificate guarantees that, from every state in $I$, the chain reaches $F_i\cup J_i\cup I^c$ at a positive time almost surely. For trajectories that remain in $I$, these conditions imply that, almost surely, either $F_i$ is visited infinitely often or $J_i$ is visited infinitely often. In the latter case, the safety bound implies that $E_i$ is visited only finitely often, almost surely. Thus, each Streett pair is satisfied almost surely on trajectories that remain in $I$.

Throughout this section, $Y$ is a DTMP
on $\mathsf S$ with transition kernel $K$, and $C\subseteq\mathsf S$
is a Borel target set. In Section~\ref{subsec:main}, $\mathsf S$ and $K$ are $\SZ$ and $K_Z$, or
$\SPsi$ and $K_\Psi$, for visit requirements, and $\Sphi$ and $K_\varphi$ for LTL.

\subsection{Reachability Certificates}
\label{subsec:reachability}

The Streett certificates include conditions for almost-sure reachability of certain target sets. To certify almost-sure reachability, the rule below uses a drift function $V$ and a variant function $U$.

\begin{definition}[Reachability Certificate]
\label{def:reachability}
A reachability certificate for $C$ is a tuple $\mathcal R=(\Inv,V,U,H,d,\epsilon)$ of a Borel set
$\Inv\supseteq C$ with $K(s,\Inv)=1$ for every $s\in\Inv\setminus C$, Borel functions
$V,U:\mathsf S\to\R_{\ge0}$, and functions $H:\R_{\ge0}\to\R_{\ge0}$, $d:\R_{\ge0}\to\R_{>0}$ and
$\epsilon:\R_{\ge0}\to(0,1]$, such that the following conditions hold:
\begin{enumerate}
  \item[(R1)] For every $s\in\Inv$: $V(s)=0$, if and only if $s\in C$.
  \item[(R2)] For every $s\in\Inv\setminus C$:
 $(KV)(s)\le V(s).$
  \item[(R3)] For every $s\in\Inv$ and every $r\ge0$ with $V(s)\le r$:
  \begin{equation*}U(s)\le H(r).\end{equation*}
  \item[(R4)] For every $s\in\Inv$: $U(s)=0$, if and only if $s\in C$.
  \item[(R5)] For every $r\ge0$ and every $s\in\Inv\setminus C$ with $U(s)\le r$:
  \begin{equation*}K\bigl(s,\{s'\in\Inv:U(s')\le U(s)-d(r)\}\bigr)\ge\epsilon(r).\end{equation*}
\end{enumerate}
\end{definition}

Since $K(s,\Inv)=1$ for every $s\in\Inv\setminus C$,
a chain started in $\Inv$ remains in $\Inv$ until it reaches $C$. By (R1), the drift $V$ vanishes
exactly on $C$, as in~\cite{labbadi2026formal}. By (R2), $V$ does not increase in expectation outside
$C$. Hence, on the event that $C$ is never reached, $V(Y_k)$ converges almost surely, and there is an
$r\ge0$ with $V(Y_k)\le r$ for infinitely many $k$. By (R3), the variant $U$ is upper bounded by
$H(r)$ on the sublevel set $\{V\le r\}$. Therefore, $U(Y_k)\le H(r)$ for the same infinitely many
$k$. By (R4), $U$ vanishes exactly on $C$. By (R5), at every state $s\in\Inv\setminus C$ with
$U(s)\le H(r)$, the variant decreases by at least $d(H(r))$ in one transition with probability at
least $\epsilon(H(r))$, and the next state belongs to $\Inv$ and satisfies $U\le H(r)$. Since $U$ is
nonnegative, at most $M(r)=\lceil H(r)/d(H(r))\rceil$ consecutive decreases make $U$ zero, and the
chain is then in $C$ by (R4). Consequently, from every state $s\in\Inv$ with $V(s)\le r$, the chain
reaches $C$ within $M(r)$ steps with probability at least $\epsilon(H(r))^{M(r)}>0$. By the Markov
property, this bound holds at each of the infinitely many times $k$ with $V(Y_k)\le r$, and it does
not depend on $k$. By L\'evy's zero-one law~\cite[Thm.~4.6.9]{durrett2019probability}, the event
that $C$ is never reached has probability zero. The following theorem states this result
and its converse.

Definition~\ref{def:reachability} is the rule of~\cite[Rule~4.1]{labbadi2026formal}, with three changes.
First, the drift condition there may fail on an exempt set, and we take this set equal to the target set.
Second, we impose (R1) on $\Inv$ only.
Third, we conclude reachability from every state of $\Inv$ instead of from one initial state that
lies in $\Inv$.
The rule of~\cite[Thm.~4.1]{mciver2018new}, stated there for programs with discrete distributions,
is the special case $U=V$ and $H(r)=r$, with nonincreasing $d$ and $\epsilon$.

\begin{theorem}[Sound and Complete Reachability Certificates]
\label{thm:reachability}
Let $C\subseteq\mathsf S$ be a Borel target set.
\renewcommand{\labelenumi}{(\alph{enumi})}%
\begin{enumerate}
    \item If $\mathcal R=(\Inv,V,U,H,d,\epsilon)$ is a reachability certificate for $C$, then $\Pr_s(\sigma_C<\infty)=1$
for every $s\in\Inv$.
\item  Let $\nu$ be a finite measure on $\mathsf S$ that gives no mass to the states from which $C$ is
reached with probability less than one, i.e.,
$  \nu\bigl(\{s\in\mathsf S:\Pr_s(\sigma_C<\infty)<1\}\bigr)=0.
$
Then there is a reachability certificate for $C$ whose set
$\Inv$ has full $\nu$-measure, i.e., $\nu(\mathsf S\setminus\Inv)=0$. In particular, if $\Pr_\mu(\sigma_C<\infty)=1$ for an initial distribution $\mu$,
such as the point mass at an initial state, there is a reachability certificate for $C$ with
$\mu(\Inv)=1$.
\item If $\Pr_s(\sigma_C<\infty)=1$ for every $s\in\mathsf S$, then for every initial distribution $\mu$ there is a
reachability certificate for $C$ such that, under $\Pr_\mu$, $Y_k\in\Inv$ for every $k$ almost
surely, also after the chain has reached $C$.
\end{enumerate}
\end{theorem}

Thus, $C$ is reached from $s$ with probability one if and only if some reachability certificate has $s\in\Inv$. The proofs are in the Appendix. For completeness, we build the variant from
layers of states that reach the target set with probability one, and the drift from the
probabilities of leaving these layers before the target set is reached.

\subsection{Streett Certificates}
\label{subsec:streett}

We now define Streett certificates. For Borel Streett pairs $(E_i,F_i)$, $i=1,\dots,\ell$, on $\mathsf S$, let $\Str$ denote their acceptance event.

\begin{definition}[Streett Certificate]
\label{def:streett}
A Streett certificate for the pairs $(E_i,F_i)$, $i=1,\dots,\ell$, consists of Borel sets $I$ and
$J_i\subseteq I\setminus E_i$, Borel functions $V_0,V_1,\dots,V_\ell:\mathsf S\to\R_{\ge0}$, constants
$\gamma_i\in(0,1]$, and reachability certificates $\mathcal R_i$, such that the following conditions
hold:
\begin{enumerate}
  \item[(S1)] The function $V_0$ satisfies
  \begin{equation*}
    \begin{aligned}
      (KV_0)(s)&\le V_0(s), &&s\in I,\\
      V_0(s)&\ge1,         &&s\notin I.
    \end{aligned}
  \end{equation*}

  \item[(S2)] For every $i=1,\dots,\ell$, the function $V_i$ satisfies
  \begin{equation*}
    \begin{aligned}
      (KV_i)(s)&\le V_i(s), &&s\notin E_i,\\
      V_i(s)&\ge1,         &&s\in E_i,\\
      V_i(s)&\le1-\gamma_i, &&s\in J_i.
    \end{aligned}
  \end{equation*}

  \item[(S3)] For every $i=1,\dots,\ell$, $\mathcal R_i$ is a reachability
  certificate for $C_i=F_i\cup J_i\cup I^c$, with an invariant set
  $\Inv_i$ containing $I$.
\end{enumerate}
\end{definition}
Thus, the invariance certificate $V_0$ bounds the probability of leaving $I$, the safety certificate $V_i$ bounds the probability of reaching the request set $E_i$ from $J_i$ by $1-\gamma_i<1$, and the reachability certificate $\mathcal R_i$ certifies that the chain reaches $C_i$ at a positive time almost surely from every state of $I$. Since $\Inv_i$ contains the target set $C_i\supseteq I^c$, condition (S3) implies $\Inv_i=\mathsf S$. Conditions (S1) and (S2) correspond to the invariance and safety conditions of~\cite[Thms.~4.1 and~4.2]{abate2026complete}. For (S3), we use the reachability certificates of Section~\ref{subsec:reachability} in place of the reachability rules of~\cite[Sec.~4]{abate2026complete}.

\begin{theorem}[Sound and Complete Streett Certificates]
\label{thm:streett}
Let $(E_i,F_i)$, $i=1,\dots,\ell$, be Borel Streett pairs on $\mathsf S$.
\begin{enumerate}
  \item[(a)] If a Streett certificate exists, then $\Pr_\mu(\Str)\ge1-\int V_0\,\dd\mu$ for every initial distribution $\mu$.
  \item[(b)] If $\Pr_\mu(\Str)=1$, there is a Streett certificate with $\int V_0\,\dd\mu=0$.
\end{enumerate}
\end{theorem}
The proofs are in the Appendix. Thus, for a fixed initial state $s$, $\Pr_s(\Str)=1$ if and only if there exists a Streett certificate with $V_0(s)=0$.

\subsection{Certificates for SDE Specifications}
\label{subsec:main}
We now establish sound and complete certificates for temporal specifications of the SDE.

\begin{theorem}[Sound and Complete Certificates for SDE Specifications]
\label{thm:main}
Consider the SDE~\eqref{eq:sde} under Assumption~\ref{ass:sde},
with initial state $X_0=x\in\R^n$.
\begin{enumerate}
  \item[(a)] Let $\Psi$ be an Emerson--Lei condition that has a conjunctive normal form with at most
  one $\Fin$ literal per clause. Then $\Pr_x(X\models\Psi)=1$ if and only if, for some
$\chi_0\in\{0,1\}^N$, the interval-hit process admits a Streett certificate for the
  pairs $(E_i,F_i)$ associated with $\Psi$
in Section~\ref{subsec:emerson-lei} with $V_0(x,\chi_0)=0$.
\item[(b)] For any other visit requirement $\Psi$, we also record
the state of the automaton for $\Psi$, as in
Lemma~\ref{lem:product-Psi}.
Then $\Pr_x(X\models\Psi)=1$ if and only if, with $\chi_0=\mathbf 0$, this process admits a Streett certificate
for the pairs $(\SZ\times E^\Psi_i,\SZ\times F^\Psi_i)$ with
$
  V_0\bigl((x,\mathbf 0),
  \Delta_\Psi(q^\Psi_{\rm in},\mathbf 0)\bigr)=0.
$
 \item[(c)] Let Assumption~\ref{ass:switching} hold, let
$h\in\Sigma(x)$ be an initial letter, and let $\varphi$
be an LTL formula. Then
$\Pr_{(x,h)}(\tr(\Lambda)\models\varphi)=1$ if and only if
the process $(S_k,q_k)$ of Theorem~\ref{thm:ltl}, which combines
the event-indexed process with the automaton for $\varphi$,
admits a Streett certificate for the pairs $(E_i,F_i)$ defined
in Section~\ref{sec:certificates}, with
$
  V_0\bigl((x,h),\Delta(q_{\rm in},h)\bigr)=0.
$
\end{enumerate}
\end{theorem}

Theorem~\ref{thm:main} covers two kinds of requirements. Visit requirements concern only which closed sets the trajectory visits, at least once or at arbitrarily large times. Parts (a) and (b) verify them on the interval-hit process, without a label, and infinitely many boundary crossings within an interval do not change the hit vectors. For example, for reachability alone, a reachability certificate for $\widetilde G_j$ with $(x,\mathbf 0)\in\Inv$ exists if and only if $\Pr_x(\sigma_{G_j}<\infty)=1$, by Corollary~\ref{cor:emerson-lei} and Theorem~\ref{thm:reachability}. The remaining requirements, for example ``$a$ until $b$'', depend on the order of visits, which the hit vectors do not record within an interval. We pose them on the hysteretic label, and part (c) verifies them, with or without \emph{next}, on its trace. Every condition
in Definitions~\ref{def:reachability} and~\ref{def:streett} involves at most one transition, whose kernel Sections~\ref{subsec:kernel-Z} and~\ref{subsec:kernel-Lambda} characterize.

\section{Numerical Examples}
\label{sec:examples}

We illustrate the certificates on four examples: recurrence to a single point, fairness, an LTL formula with the \emph{next} operator, and an ordered LTL task for a nonlinear planar SDE. The certificates of the first three examples are in closed form. The certificate of the fourth is defined through PDE solutions, which we compute numerically.

\subsection{Recurrence Verification}
\label{subsec:example-recurrence}

The first example certifies recurrence. Let $n=m=1$ and $\dd X_t=-X_t\,\dd t+\dd W_t$, an Ornstein--Uhlenbeck (OU) process,
which satisfies Assumption~\ref{ass:sde} with $c_0=1$. The target set is the single point $G=\{0\}$, which the samples $X_{k\delta}$ alone miss, since $X_{k\delta}\ne0$ almost surely for every $k\ge1$,
and we set the sampling period $\delta=1/2$. Given $X_0=x$, the state $X_\delta$ is normal with mean $\alpha x$ and variance
$v$, where $\alpha=e^{-1/2}$ and $v=(1-e^{-1})/2$. We certify that $\Pr_x(\Inf(\{0\}))=1$ for every
$x\in\R$. Here, $N=1$ and $G_1=G$. The interval-hit process is $Z_k=(X_{k\delta},\chi_k)$ with
state space $\SZ=\R\times\{0,1\}$, where, for $k\ge1$, $\chi_k=1$ if the trajectory visits $G$ during $[(k-1)\delta,k\delta]$ and $\chi_k=0$ otherwise. Thus, $\widetilde G=\R\times\{1\}$ consists of the states that record a visit to $G$ in the preceding interval.

Recurrence of $G$ is the Streett condition with the single pair $(\SZ,\widetilde G)$ of
Section~\ref{subsec:emerson-lei}. Definitions~\ref{def:reachability} and~\ref{def:streett} are
applied with $\mathsf S=\SZ$ and $K=K_Z$. We take $I=\SZ$, $J_1=\varnothing$, $V_0=0$, $V_1=1$ and
$\gamma_1=1$. They satisfy (S1), since $V_0=0$, and (S2), since $V_1=1$ on the request set $\SZ$ and the absorbing region is empty. The target set of (S3) is $C_1=\widetilde G$.
We show that the following choices define a reachability certificate $\mathcal R_1$ for $C_1$:
$\Inv_1=\SZ$, $H(r)=r$, $d(r)=1/2$, $\epsilon(r)=0.28$, $V(x,\eta)=U(x,\eta)=0$ if $\eta=1$, and $V(x,\eta)=U(x,\eta)=1+x^2$ if $\eta=0$.
Conditions (R1), (R3) and (R4) hold by construction, and (R2) and (R5) are required only at
the states $(x,0)$. By Lemma~\ref{lem:inclusion-exclusion} for the single target set $G=\{0\}$,
$K^0(x,\dd y)=p^1_\delta(x,y)\,\dd y$ and
$K^1(x,\dd y)=\bigl(p_\delta(x,y)-p^1_\delta(x,y)\bigr)\dd y$, where
$p_\delta(x,y)=(2\pi v)^{-1/2}\exp\bigl(-(y-\alpha x)^2/(2v)\bigr)$ is the Gaussian transition density
and $p^1_\delta$ is the density of the transitions without a visit to $G$
(Proposition~\ref{prop:killed-parabolic}(a), applicable since $g=1$). By the strong Markov property at
the first visit to zero and the symmetry of the SDE under $x\mapsto-x$, the trajectories that visit zero
have equal densities at $y$ and $-y$. For $xy>0$, every trajectory ending at $-y$ visits zero, and hence $p^1_\delta(x,y)=p_\delta(x,y)-p_\delta(x,-y)$. For $xy\le0$, every trajectory from $x$ to $y$ visits zero, and hence $p^1_\delta(x,y)=0$.

For (R2), by~\eqref{eq:kernel-Z} and $V(y,1)=0$, direct integration, with the symmetry $p_\delta(-x,-y)=p_\delta(x,y)$, the substitutions
$y=\pm\alpha\lvert x\rvert+\sqrt v\,\theta$ and an integration by parts, gives
\begin{align*}
  &(K_ZV)(x,0)=\int_\R(1+y^2)\,p^1_\delta(x,y)\,\dd y\\*
  &=(1+\alpha^2x^2+v)\Bigl(2\Phi\Bigl(\frac{\alpha\lvert x\rvert}{\sqrt v}\Bigr)-1\Bigr)+2\alpha\sqrt v\lvert x\rvert\Phi'\Bigl(\frac{\alpha\lvert x\rvert}{\sqrt v}\Bigr),
\end{align*}
where $\Phi$ is the standard normal distribution function.

\begin{figure}[t]
  \centering
  \includegraphics[scale=1]{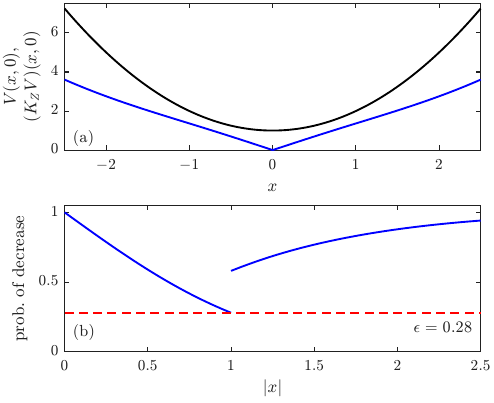}
  \caption{Recurrence example. (a) $V(x,0)=1+x^2$ (black) and $(K_ZV)(x,0)$ (blue). (b) Lower bound (blue) on the probability that $U$ decreases by at least $1/2$, and the level $\epsilon=0.28$ (red, dashed).}
  \label{fig:ou-certificate}
\end{figure}

Since $p^1_\delta\le p_\delta$, the second moment of $X_\delta$ gives $(K_ZV)(x,0)\le1+\alpha^2x^2+v\le1+x^2$ for $\lvert x\rvert\ge1$, by $\alpha^2+2v=1$. For $\lvert x\rvert<1$, the bounds $2\Phi(\theta)-1\le2\theta/\sqrt{2\pi}$ for $\theta\ge0$, $\Phi'(\theta)\le1/\sqrt{2\pi}$ and $1+\alpha^2x^2+2v\le2$ give $(K_ZV)(x,0)\le4\lvert x\rvert/\sqrt{\pi(e-1)}\le2\lvert x\rvert\le1+x^2$. Thus, $(K_ZV)(x,0)\le V(x,0)$, which is (R2), as Fig.~\ref{fig:ou-certificate}(a) shows.

For (R5), it suffices to show that $\Pr_{(x,0)}\bigl(U(Z_1)\le U(x,0)-1/2\bigr)\ge0.28$ for every
$x\in\R$. A visit to $G$ during $[0,\delta]$ gives $\chi_1=1$ and $U(Z_1)=0\le U(x,0)-1/2$, since
$U(x,0)\ge1$, and it has probability $K^1(x,\R)=1-\int_\R p^1_\delta(x,y)\,\dd y=2\Phi\bigl(-\alpha\lvert x\rvert/\sqrt v\bigr)$. For
$\lvert x\rvert\le1$, the monotonicity of $\Phi$ gives
$K^1(x,\R)\ge2\Phi(-\alpha/\sqrt v)>0.28$. For $\lvert x\rvert>1$, let $\beta=\sqrt{x^2-1/2}$. The
required decrease occurs exactly when the trajectory visits $G$, or does not visit $G$ and ends in
$[-\beta,\beta]$, where $1+X_\delta^2\le1+x^2-1/2$. These disjoint events give
\begin{align*}
  K^1\!(x,\!\R)\!+\!K^0\!(x,\![-\beta,\beta])
  \!\!=\!\Phi\!\Bigl(\!\frac{\beta-\alpha\lvert x\rvert}{\sqrt v}\!\Bigr)\!+\!\Phi\!\Bigl(\!\frac{-\beta-\alpha\lvert x\rvert}{\sqrt v}\!\Bigr)\!\!>\!\!\frac12,
\end{align*}
since $\beta^2-\alpha^2x^2=(1-e^{-1})x^2-1/2>0$. This verifies (R5), and Fig.~\ref{fig:ou-certificate}(b)
shows the resulting lower bound. Hence, $\mathcal R_1$ gives (S3), and
Theorem~\ref{thm:main}(a) yields the claim.

\subsection{Fairness Verification}
\label{subsec:example-fairness}
The second example verifies a fairness requirement. Let $n=m=1$ and $\dd X_t=\dd t+\dd W_t$, which satisfies Assumption~\ref{ass:sde} with $c_0=2$, and
let the sampling period $\delta=1$. The requirement is strong fairness, $\Inf(E)\Rightarrow\Inf(F)$, with the request set
$E=(-\infty,0]$ and a closed response set $F$. The trajectory tends to $+\infty$ almost surely. Hence,
almost surely, $X_t\notin E$ for all sufficiently large $t$, and the requirement holds for every $F$.
When $F$ is bounded, $X_t\notin F$ also holds for all sufficiently large $t$. A certificate must then use a non-empty absorbing region $J_1$, since with $J_1=\varnothing$ it
would certify that $F$ is visited at arbitrarily large times, which is false.

The interval-hit process records the two target sets $G_1=E$ and $G_2=F$, its states are $(x,\eta)$ with
$\eta=(\eta_1,\eta_2)$ and state space $\SZ=\R\times\{0,1\}^2$. The Streett pair is $(\widetilde E,\widetilde F)$, where $ \widetilde E=\{(x,\eta)\in\SZ:\eta_1=1\}$ and $\widetilde F=\{(x,\eta)\in\SZ:\eta_2=1\}$. We take $I=\SZ$ and
$V_0=0$, which satisfy (S1).

To verify (S2), we choose the absorbing region $J_1=\{(x,\eta):x\ge1,\ \eta_1=0\}$ and
$V_1(x,\eta)=\Pr_s(\sigma_{\widetilde E}<\infty)$ with $s=(x,\eta)$, which equals $1$ if $\eta_1=1$ or $x\le0$, and $e^{-2x}$ if $\eta_1=0$ and $x>0$.
If $\eta_1=1$, the state already belongs
to $\widetilde E$. Otherwise, $Z$ reaches $\widetilde E$
exactly when $X$ visits $E$, which is certain for $x\le0$.
For $x>0$ and $\bar x>x$, the probability $u(x)$ that $X$ reaches zero before $\bar x$ solves
$u'+\frac12u''=0$ on $(0,\bar x)$ with $u(0)=1$ and $u(\bar x)=0$, and equals
$(e^{-2x}-e^{-2\bar x})/(1-e^{-2\bar x})$. Letting $\bar x\to\infty$ gives $\Pr_x(\sigma_E<\infty)=e^{-2x}$,
which establishes the stated expression for $V_1$.  For $s\notin\widetilde E$, $\sigma_{\widetilde E}\ge1$, and the Markov property at step one gives
$(K_ZV_1)(s)=\E_s[V_1(Z_1)]=\Pr_s(\sigma_{\widetilde E}<\infty)=V_1(s)$.
Moreover, $V_1=1$ on $\widetilde E$ and
$V_1\le e^{-2}<0.14$ on $J_1$, since $x\ge1$ there.
Hence, (S2) holds with $\gamma_1=0.86$.

The reachability certificate $\mathcal R_1$ for $C_1=\widetilde F\cup J_1$ uses $\Inv_1=\SZ$, the
drift and the variant
\begin{align*}
  V(x,\eta)&=
  \begin{cases}
    0, & (x,\eta)\in C_1,\\
    e^{-x}, & (x,\eta)\notin C_1,
  \end{cases}\\
  U(x,\eta)&=
  \begin{cases}
    0, & (x,\eta)\in C_1,\\
    1+\max\{1-x,0\}, & (x,\eta)\notin C_1,
  \end{cases}
\end{align*}
and $d=1/2$,
$\epsilon=0.38$, $H(r)=1+\max\{1+\ln r,0\}$ for $r>0$ and $H(0)=1$. Conditions (R1) and (R4) hold,
since $V>0$ and $U\ge1$ on $\SZ\setminus C_1$. Condition (R2) holds, since $V(y,\eta)\le e^{-y}$ for every state
and $\E_x[e^{-X_\delta}]=e^{-x-\delta/2}\le e^{-x}$. Condition (R3) holds, since, for $(x,\eta)\notin C_1$, $V(x,\eta)\le r$
forces $r>0$ and $x\ge-\ln r$, and hence $U\le H(r)$, while $U=0$ on $C_1$. For (R5), a state with
$x\le1/2$ has $X_\delta\ge x+1/2$ with probability $\Phi(1/2)\ge0.69$, and then $U$ decreases by
at least $1/2$ or the chain enters $C_1$. A state $(x,\eta)\notin C_1$ with $x>1/2$ has $U\ge1$, and $U$
decreases by at least $1/2$ when the chain enters $C_1$, in particular when the chain ends in $J_1$. On the event that $W$ stays above $-x$ on $[0,\delta]$, $X_t=x+t+W_t>t\ge0$, so $X$ does not visit $E$ and $X_\delta>\delta=1$, that is, the chain ends in $J_1$. By the reflection principle~\cite[eq.~(2.6.2)]{karatzas1991brownian}
and the symmetry of $W$, this event has probability $2\Phi(x/\sqrt\delta)-1$, which increases with $x$ and is at least
$2\Phi(1/2)-1\ge0.38$. These sets and functions form a Streett certificate, and
Theorem~\ref{thm:main}(a) gives $\Pr_x(\Inf(E)\Rightarrow\Inf(F))=1$ for every $x$.
\subsection{LTL Verification: Alternating Label}
\label{subsec:example-ltl}

The third example uses the \emph{next} operator. Let $X$ be the OU process of the first
example, $\AP=\{a\}$, and $A_a=[0,\infty)$. The memoryless label, $L(x)=\{a\}$ for $x\ge0$ and
$L(x)=\varnothing$ for $x<0$, changes infinitely often immediately after the first visit of the trajectory to $0$, which
occurs almost surely by the first example. The chattering follows as in
Example~\ref{ex:chattering}. The hysteretic label of Lemma~\ref{lem:band} with $\rho_a=1$ has
$A^1_a=[1,\infty)$ and $A^0_a=(-\infty,-1]$. We verify the LTL formula
\begin{equation*}
\varphi=\Box(a\Rightarrow\Snext\neg a)
  \wedge\Box(\neg a\Rightarrow\Snext a).
\end{equation*}
It requires the trace to alternate between $a$ and $\neg a$. If switching stops, the trace repeats its last letter forever,
violating $\varphi$. As an event of $X$, $\varphi$ is the visit requirement $\Inf(A^1_a)\wedge\Inf(A^0_a)$, and we verify it here with \emph{next} to illustrate Theorem~\ref{thm:main}(c).

From a state $(x,\{a\})$ the next switch occurs at the first visit to $-1$, and from $(x,\varnothing)$ at
the first visit to $1$. For the OU process, every point is reached almost surely
from every initial state~\cite[Prop.~5.5.22(a)]{karatzas1991brownian}.
Thus, from every consistent state, the event-indexed process
moves from $(x,\{a\})$ to $(-1,\varnothing)$ and from
$(x,\varnothing)$ to $(1,\{a\})$ with probability one.

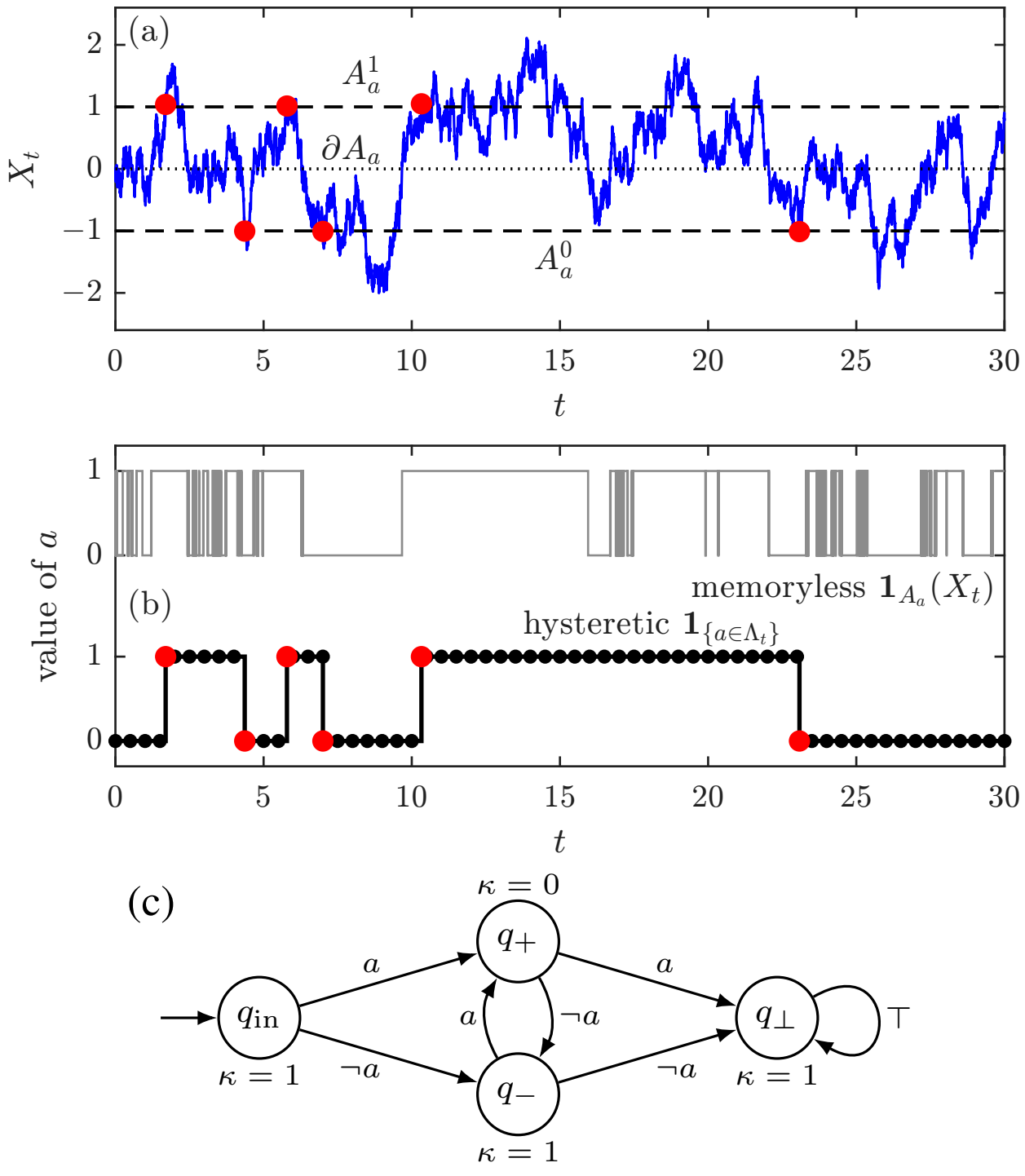
\begin{figure}[t]
  \centering
  \includegraphics[scale=1]{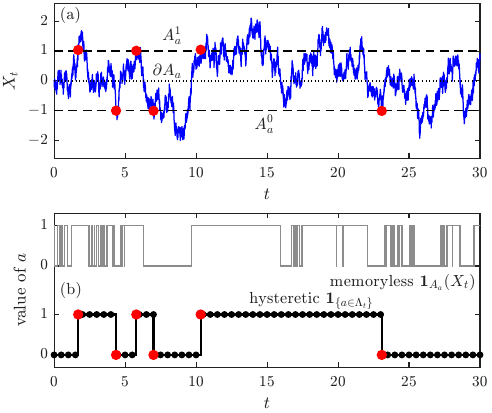}\\[2pt]
  \begin{tikzpicture}[font=\footnotesize,>={Latex[length=1.6mm]},semithick,
      st/.style={draw,circle,minimum size=6.6mm,inner sep=0pt},
      pr/.style={font=\scriptsize,inner sep=1pt},
      lab/.style={font=\scriptsize,inner sep=1.5pt}]
    \node[st] (in) at (0,0) {$q_{\rm in}$};
    \node[st] (p) at (2.1,0.62) {$q_+$};
    \node[st] (m) at (2.1,-0.62) {$q_-$};
    \node[st] (b) at (4.2,0) {$q_\perp$};
    \node[pr,below=0pt of in] {$\kappa=1$};
    \node[pr,above=0pt of p] {$\kappa=0$};
    \node[pr,below=0pt of m] {$\kappa=1$};
    \node[pr,below=0pt of b] {$\kappa=1$};
    \draw[->] (-0.8,0) -- (in);
    \draw[->] (in) -- node[lab,above left] {$a$} (p);
    \draw[->] (in) -- node[lab,below left] {$\neg a$} (m);
    \draw[->] (p) to[bend left=30] node[lab,right] {$\neg a$} (m);
    \draw[->] (m) to[bend left=30] node[lab,left] {$a$} (p);
    \draw[->] (p) -- node[lab,above right] {$a$} (b);
    \draw[->] (m) -- node[lab,below right] {$\neg a$} (b);
    \draw[->] (b) to[out=30,in=-30,looseness=6] node[lab,right] {$\top$} (b);
    \node[anchor=north west,font=\normalsize] at (-1.2,1.2) {(c)};
  \end{tikzpicture}
  \caption{(a) A trajectory of the OU process of the LTL example from $X_0=0$ and $\Lambda_0=\varnothing$ (blue), the boundaries
  of $A^1_a$ and $A^0_a$ (black, dashed) and $\partial A_a=\{0\}$ (black, dotted). (b) The memoryless label $\one_{A_a}(X_t)$ (gray), the hysteretic label (black) and its samples at $k\delta$, $\delta=1/2$ (black dots), where successive samples repeat a letter. Red dots in (a) and (b) mark the switches of the hysteretic label, each at the first grid point in the switching set reached. (c) The parity automaton for $\varphi$: edge labels give the truth value of $a$ in the letter read ($\top$ for either letter), and $\kappa$ is the priority of a state.}
  \label{fig:ou-hysteresis}
\end{figure}

Fix a sampling period $\delta>0$. After the first switch, the successive
holding times of the hysteretic label are independent and identically
distributed by the strong Markov property and the symmetry of the OU
process. Each holding time exceeds $2\delta$ with positive probability.
Since switching continues almost surely, some holding
interval exceeds $2\delta$ almost surely and therefore contains two consecutive
sampling instants. The sampled word $(\Lambda_{k\delta})_{k\in\N}$
thus contains two consecutive identical letters and violates $\varphi$
almost surely. Fig.~\ref{fig:ou-hysteresis}(a),(b) illustrates this for $\delta=1/2$.

The automaton for $\varphi$ is shown in
Fig.~\ref{fig:ou-hysteresis}(c) and starts in $q_{\rm in}$. The states $q_+$ and $q_-$ record that the last letter was
$\{a\}$ or $\varnothing$, respectively. A repeated letter violates $\varphi$ and leads to $q_\perp$, where the automaton remains forever. With priority zero at $q_+$ and priority one at all other states, a run is accepted exactly
when it visits $q_+$ infinitely often. The Streett condition on the process
$(S_k,q_k)$ therefore has the single pair $E_1=\SL\times\{q_{\rm in},q_-,q_\perp\}$ and $F_1=\SL\times\{q_+\}$.

Let $I$ be the set of product states $((y,h),q)$ with $y\in\R$, $q=q_+$ if $a\in h$, and $q=q_-$
otherwise. It contains every initial state $((x,h),\Delta(q_{\rm in},h))$, and the product process stays
in $I$ with probability one. The Streett certificate takes $V_0=\one_{I^c}$, $J_1=\varnothing$, $V_1=1$
and $\gamma_1=1$. Thus, (S1) holds, with $V_0=0$ at every initial state, and (S2) holds. The target set is $C_1=F_1\cup I^c$, whose complement consists of the states of $I$ with
$q=q_-$. The reachability certificate $\mathcal R_1$ for $C_1$ has $\Inv_1=\Sphi$,
$V=U=\one_{C_1^c}$, $H=1$, $d=1$ and $\epsilon=1$, since from a state of $C_1^c$, where the label is $\varnothing$, the next switch is the visit to $1$, which occurs with probability one, and after it $q=q_+$. Theorem~\ref{thm:main}(c) gives $\Pr_{(x,h)}(\tr(\Lambda)\models\varphi)=1$ for every
consistent pair $(x,h)$.

\begin{figure*}[!t]
  \centering
  \includegraphics[scale=1]{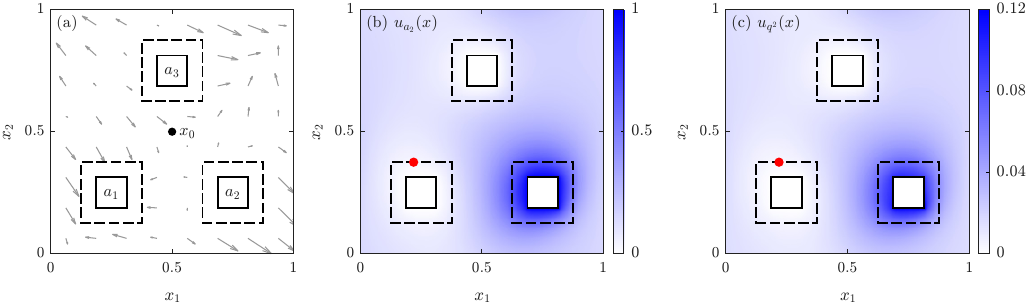}
  \caption{Planar example. (a) Boundaries of $A^1_a$ (solid) and $A^0_a$ (dashed), the drift $f$ (gray arrows) and $x_0$ (black dot). (b), (c) Probabilities $u_{a_2}$ and $u_{q^2}$ outside the activation squares, with the minimizers over $\Upsilon$ attaining $\epsilon$ and $\gamma_1$ (red dots).}
  \label{fig:pde-example}
\end{figure*}

\subsection{LTL Verification: Ordered Task with PDE Kernels}
\label{subsec:example-pde}

We verify an ordered LTL task for a nonlinear
planar SDE. Let $n=m=2$ and $\dd X_t=f(X_t)\,\dd t+0.35\,\dd W_t$, with $X_0=x_0=(1/2,1/2)$ and
\begin{align*}
  f_1(x)&=0.15+0.35\sin2\pi(x_2-x_1)-0.25\sin2\pi x_1,\\
  f_2(x)&=-0.10+0.30\sin2\pi(x_1-x_2)-0.20\sin2\pi x_2.
\end{align*}
Since $f$ is smooth and bounded, Assumption~\ref{ass:sde} holds.
Its $1$-periodicity allows us to represent the process in the unit square,
where a trajectory leaving one side continues from the opposite side.
This periodic space is denoted by $\mathbb T^2$.
The diffusion is uniformly elliptic since $gg^\top=0.1225I_2$. Let
$\AP=\{a_1,a_2,a_3\}$. For $a\in\AP$, the activation set $A^1_a$ is a closed square of half-width
$1/16$, and the deactivation set $A^0_a$ is the complement of the open square of half-width $1/8$ with
the same center. The centers are $(1/4,1/4)$, $(3/4,1/4)$ and $(1/2,3/4)$ for $a_1$, $a_2$ and $a_3$,
respectively, and all sets are repeated with period $1$ in each coordinate, so that
Assumption~\ref{ass:switching} holds (Fig.~\ref{fig:pde-example}(a)). Since the squares of half-width
$1/8$ are disjoint, a trajectory leaves the square of an active proposition before reaching another activation set. Hence, the label alternates between $\varnothing$ and $\{a_1\}$, $\{a_2\}$ or $\{a_3\}$. Since $x_0$ lies in no activation
set, we take $\Lambda_0=\varnothing$ and verify
\begin{equation*}
  \begin{aligned}
    \varphi&=\Diamond\bigl(\{a_1\}\wedge\Snext(\varnothing\wedge\Snext(\{a_2\}\wedge
      \Snext(\varnothing\wedge\Snext\{a_3\})))\bigr)\\
    &\quad\wedge\Box\Diamond\bigl(\{a_3\}\wedge\Snext\varnothing\bigr),
  \end{aligned}
\end{equation*}
where letters denote formulas as in Remark~\ref{rem:next}. The
formula requires the word $\{a_1\}\varnothing\{a_2\}\varnothing\{a_3\}$ in the trace and infinitely many
occurrences of $\{a_3\}\varnothing$.

The deterministic parity automaton $\mathcal A_\varphi$ has the states
$q_{\rm in}=q^0,q^1,\dots,q^7$. For every $j=0,\dots,4$, the state $q^j$ records that the longest suffix
of the letters read that is a prefix of the word has length $j$. On the next letter of the word, it
moves to $q^{j+1}$, except that $q^4$ moves to $q^6$. On any other letter, it moves to $q^1$ if the
letter is $\{a_1\}$ and to $q_{\rm in}$ otherwise. After the first match, $q^6$ records that the last
letter is $\{a_3\}$, $q^7$ that the last two letters are $\{a_3\}\varnothing$, and $q^5$ the remaining
case. The priority $\kappa$ equals $1$ on $\{q_{\rm in},q^1\}$, $4$ on $\{q^2,q^3,q^4\}$, $3$ on
$\{q^5,q^6\}$ and $2$ on $\{q^7\}$. The Streett pairs of Section~\ref{sec:certificates} are
$E_1=\SL\times\{q_{\rm in},q^1\}$, $F_1=\varnothing$, $E_2=\SL\times\{q^5,q^6\}$ and
$F_2=\SL\times\{q^7\}$.

From $(x,\varnothing)$, the next switch is the first visit to $A^1_{a_1}\cup A^1_{a_2}\cup A^1_{a_3}$, and from $(x,\{a\})$ the first visit to $A^0_a$. On the compact torus, the uniformly elliptic diffusion reaches every set with nonempty interior almost surely, so these switches occur almost surely, and every boundary point of the squares is regular for $X$, since the switching set contains a cone with that vertex. Moreover, since the squares are disjoint, $\Gamma(h,\cdot)$ is constant near every boundary point and the boundary data below are continuous, so that $\hat b$ is continuous. Hence, Proposition~\ref{prop:exit-elliptic}(b) applies, with exit problems on $\mathbb T^2$. Let $u_a(x)$ be the probability that the
next activated proposition is $a$. The function $u_a$ solves $\mathcal Lu_a=0$ outside
$A^1_{a_1}\cup A^1_{a_2}\cup A^1_{a_3}$, with $u_a=1$ on $\partial A^1_a$ and $u_a=0$ on the boundaries
of the other two activation sets. For $j=2,3,4$, let $u_{q^j}(y)$ be the probability that the word is
completed before $E_1$ is visited, from a state $((y,h),q^j)$ immediately after a switch. Then
$u_{q^4}=u_{a_3}$, the function $u_{q^3}$ is $\mathcal L$-harmonic in the open square of $a_2$ with
boundary values $u_{a_3}$, and $u_{q^2}$ is $\mathcal L$-harmonic outside the activation sets, with
boundary values $u_{q^3}$ on $\partial A^1_{a_2}$ and $0$ on $\partial A^1_{a_1}\cup\partial A^1_{a_3}$.

With $\partial A^0=\partial A^0_{a_1}\cup\partial A^0_{a_2}\cup\partial A^0_{a_3}$, let $I$ consist of the
initial state $s_0=((x_0,\varnothing),q_{\rm in})$ of the product process and the states $((y,h),q)$ with
$y\in\partial A^1_a$ if $h=\{a\}$ and $y\in\partial A^0$ if $h=\varnothing$, where $(h,q)$ ranges over
the twelve pairs reachable from $(\varnothing,q_{\rm in})$ along alternating letters. After every switch
the position lies on these boundaries, and hence $K_\varphi(s,I)=1$ for every $s\in I$. Thus,
$V_0=\one_{I^c}$ satisfies (S1) with $V_0(s_0)=0$. Let $J_1=I\setminus E_1$, $J_2=\varnothing$, $V_2=1$,
$\gamma_2=1$ and $V_1(s)=\Pr_s(\sigma_{E_1}<\infty)$. By the Markov property,
$(K_\varphi V_1)(s)=V_1(s)$ for every $s\notin E_1$, and $V_1=1$ on $E_1$. Moreover,
$V_1=1-u_{q^j}(y)$ on $J_1$ at $q^j$, $j=2,3,4$, and $V_1=0$ at $q^5,q^6,q^7$, from which $E_1$ is
unreachable. With $\Upsilon=\{x_0\}\cup\partial A^0$, let
\begin{equation*}
  \gamma_1=\min\bigl\{\min_{\Upsilon}u_{q^2},\,\min_{\partial A^1_{a_2}}u_{q^3},\,
    \min_{\Upsilon}u_{q^4}\bigr\}
\end{equation*}
and $\epsilon=\min_{a\in\AP}\min_{\Upsilon}u_a$. By the strong maximum principle, the functions $u_a$, $u_{q^2}$, $u_{q^3}$ and $u_{q^4}$ are positive on their connected domains, and, on $\mathbb T^2$, $\Upsilon$ and $\partial A^1_{a_2}$ are compact subsets of these domains. Hence, $\gamma_1,\epsilon>0$, $V_1\le1-\gamma_1$ on $J_1$, and (S2) holds.

For (S3), let $C_1=J_1\cup I^c$ and $C_2=F_2\cup I^c$. For $i=1,2$, the reachability certificate
$\mathcal R_i$ for $C_i$ has $\Inv_i=\Sphi$, $V=\one_{C_i^c}$, $d=1$ and the constant $\epsilon$
defined above. Here $U_i(s)$ is the least number of switches that lead from $s$ to $C_i$ with positive
probability, with $U_i=0$ on $C_i$, and $H_i=\max U_i$ equals $3$ for $i=1$ and $7$ for $i=2$. Then
(R1)--(R4) hold, since $V=1\ge K_\varphi V$ on $C_i^c$, $U_i\le H_i$, and $U_i$ vanishes exactly on
$C_i$. For (R5), each of the $U_i(s)$ switches on a path to $C_i$ is a deactivation, with probability
one, or the activation of a prescribed proposition from a position in $\Upsilon$, with probability at
least $\epsilon$, which is (R5). Hence, Theorem~\ref{thm:main}(c) gives
$\Pr_{(x_0,\varnothing)}(\tr(\Lambda)\models\varphi)=1$.

We solve these problems by monotone finite differences on a $256\times256$ periodic grid and obtain
$\epsilon\approx0.038$ and $\gamma_1\approx0.004$. Fig.~\ref{fig:pde-example}(b),(c) show $u_{a_2}$ and $u_{q^2}$, whose minima over $\Upsilon$ are the smallest ones and define $\epsilon$ and $\gamma_1$. The conclusion uses only $\gamma_1,\epsilon>0$.

\section{Conclusion}
\label{sec:conclusion}

We certify that visit requirements, such as recurrence and fairness, and LTL requirements of SDEs hold with probability one, by
reducing them, on every trajectory, to acceptance conditions of discrete-time Markov processes. For
requirements on visits to closed target sets, we record which target sets the trajectory visits during each sampling interval, and the
reduction holds for every sampling period and for target sets of measure zero. For LTL, we replace the
memoryless label, which can change infinitely often at a boundary across which the noise acts, by a hysteretic label, and sample
the SDE at the switches of this label. Every formula, the \emph{next} operator included, then becomes a parity condition. On
these processes, whose state spaces are uncountable, Streett certificates built from a reachability rule with a drift and a variant function are sound and, for each initial state, complete. For a uniformly elliptic diffusion, parabolic problems characterize the kernel of the interval-hit process (Proposition~\ref{prop:killed-parabolic}), and, under the conditions of Proposition~\ref{prop:exit-elliptic}, elliptic exit problems characterize the kernel of the event-indexed process.

The requirements constrain the times of visits rather than the time spent, and the hysteretic semantics depends on the
switching sets that the designer chooses. Completeness concerns
certificates over all measurable functions. A search over a fixed class of functions remains sound but
may fail, and numerical solutions of the PDEs certify only with one-sided error bounds. Controlled
diffusions, certified numerical schemes for the killed and exit problems, and requirements on
occupation times are directions for future work.

\appendix
\label{app:proofs}

This appendix includes the proofs of all results of Sections~\ref{sec:recurrence}
to~\ref{sec:certificates} and a
lemma on
enlarging the target set of a reachability certificate, which the proof of
Theorem~\ref{thm:streett}(b) uses.

\subsection{Proofs of Section~\ref{sec:recurrence}}
\begin{IEEEproof}[Proof of Lemma~\ref{lem:markov-Z}]
As in Section~\ref{subsec:specifications}, $\chi_{k+1}$ is a Borel function of the path on
$[k\delta,(k+1)\delta]$, since the target sets are closed and the paths are continuous. The solution is a measurable function
of $x$ and $W$ (Section~\ref{subsec:model}), and $K^\eta(x,A)$ is therefore Borel in $x$. Thus,
$K^\eta$ is a subprobability kernel. The Markov property of $X$ at the time $k\delta$ gives that the conditional law of
$(X_{(k+1)\delta},\chi_{k+1})$ given $\mathcal F_{k\delta}$ is $K_Z(Z_k,\cdot)$.
\end{IEEEproof}

\begin{IEEEproof}[Proof of Theorem~\ref{thm:recurrence}]
If the trajectory visits $G_j$ at arbitrarily large times, then for every $k_0\in\N$ some $t\ge k_0\delta$ with $X_t\in G_j$ lies in an interval $[k\delta,(k+1)\delta]$ with $k\ge k_0$, so $Z_{k+1}\in\widetilde G_j$ by~\eqref{eq:hit-vector}, and hence $Z_k\in\widetilde G_j$ for infinitely many $k$. Conversely, if $Z_{k_i}\in\widetilde G_j$ for an increasing sequence $k_i\ge1$, then by~\eqref{eq:hit-vector} there are $t_i\in[(k_i-1)\delta,k_i\delta]$ with $X_{t_i}\in G_j$ and $t_i\to\infty$. The event $\Inf_{\widetilde G_j}$ does not depend on the initial hit vector, and the equality of events gives~\eqref{eq:recurrence-probability}, whose right side is a probability of the DTMP with kernel $K_Z$ by Lemma~\ref{lem:markov-Z}.
\end{IEEEproof}

\begin{IEEEproof}[Proof of Corollary~\ref{cor:emerson-lei}]
Theorem~\ref{thm:recurrence} gives the equivalence for every $\Inf(G_j)$, and by complement for
every $\Fin(G_j)$, for every $\chi_0$. The intervals in~\eqref{eq:hit-vector} cover $\R_{\ge0}$, so $\sigma_{G_j}<\infty$ if and only if $\chi_{k,j}=1$ for some $k\ge1$, i.e., with $\chi_0=\mathbf 0$, if and only if $\sigma_{\widetilde G_j}<\infty$. Induction over the Boolean structure of $\Psi$ gives the claim.
\end{IEEEproof}

\begin{IEEEproof}[Proof sketch of Lemma~\ref{lem:product-Psi}]
The automaton coordinate is a deterministic function of the previous one and of the next hit
vector, and hence the Markov property of $Z$ carries over. The run of $\mathcal A_\Psi$ on
$\chi_0\chi_1\cdots$ is $q^\Psi_{\rm in},q_0,q_1,\dots$, and its acceptance depends only on the
states visited infinitely often.
\end{IEEEproof}

\begin{IEEEproof}[Proof of Lemma~\ref{lem:inclusion-exclusion}]
The event $\{\chi_1=\eta\}$ is the intersection of the events $\{\sigma_{G_j}\le\delta\}$ for
$\eta_j=1$ and $\{\sigma_{G_j}>\delta\}$ for $\eta_j=0$. Expanding
$\one_{\{\sigma_{G_j}\le\delta\}}=1-\one_{\{\sigma_{G_j}>\delta\}}$ gives $\one_{\{\chi_1=\eta\}}=\sum_{\zeta\le\eta}(-1)^{\lvert\zeta\rvert}\one_{\{\sigma_{G_{\eta^c+\zeta}}>\delta\}}$,
because no target set in $G_{\eta^c+\zeta}$ is visited exactly when $G_{\eta^c+\zeta}$ is not
hit. Multiplying by $\one_A(X_\delta)$ and taking expectations gives~\eqref{eq:inclusion-exclusion}.
\end{IEEEproof}

We first prove a localization lemma.

\begin{lemma}[Localization]
\label{lem:localization}
Let Assumption~\ref{ass:sde} hold and let the diffusion be uniformly elliptic with constant $\lambda$. For every closed ball, up to its exit time from the ball, $X$ coincides with
the solution of an SDE driven by an $n$-dimensional Brownian motion with respect to
$(\mathcal F_t)$, whose coefficients are bounded, globally Lipschitz and uniformly elliptic with the
same $\lambda$.
\end{lemma}

\begin{IEEEproof}
Since every eigenvalue of $gg^\top$ is at least $\lambda$, the matrix $(gg^\top)^{-1/2}g$ is continuous in $x$ and $(gg^\top)^{-1/2}g\,g^\top(gg^\top)^{-1/2}=I$. Hence $\int_0^t(gg^\top)^{-1/2}(X_{t_2})\,g(X_{t_2})\,\dd W_{t_2}$ is a continuous martingale with respect to $(\mathcal F_t)$ with bounded integrand and cross-variation matrix $tI$, i.e., an $n$-dimensional Brownian motion by L\'evy's characterization~\cite[Thm.~3.3.16]{karatzas1991brownian}, and $X$ solves the SDE with the coefficients $f$ and $(gg^\top)^{1/2}$ driven by it. Composing these coefficients with the radial projection onto a closed ball gives bounded, globally Lipschitz coefficients, since the matrix square root is Lipschitz on the symmetric matrices whose eigenvalues are at least $\lambda$. They are uniformly elliptic with the same $\lambda$, and the solution of the SDE with these coefficients, driven by the same Brownian motion from the same initial state, coincides with $X$ up to its exit time from that ball.
\end{IEEEproof}

\begin{IEEEproof}[Proof of Proposition~\ref{prop:killed-parabolic}]
(a) Fix $x\in\R^n$ and $t>0$. By Lemma~\ref{lem:localization}, up to the exit from a closed ball containing $x$, $X$ coincides with a diffusion with bounded, globally Lipschitz and uniformly elliptic coefficients, which has a transition density~\cite[Ch.~6, Thm.~5.4]{friedman1975stochastic}. For a Lebesgue-null set $A$, $\Pr_x(X_t\in A)$ is therefore at most the probability that $X$ leaves the ball by time $t$, which tends to zero as the radius grows, since $X$ does not explode. Hence, the law of $X_t$ has a density $p_t(x,\cdot)$. Since $\Pr_x(X_t\in A,\ \sigma_{G_\zeta}>t)\le\Pr_x(X_t\in A)$ for every Borel set $A$, the killed measure has a density $p^\zeta_t(x,\cdot)\le p_t(x,\cdot)$ almost everywhere.

(b) By Lemma~\ref{lem:localization} with a closed ball containing $\overline D$, $X$ coincides up to its exit from $D$ with a diffusion driven by an $n$-dimensional Brownian motion whose coefficients are bounded, globally Lipschitz and uniformly elliptic. Since $b=0$ on $\partial D$, the initial and boundary data agree there, and the parabolic representation in~\cite[Ch.~6, Thm.~5.2]{friedman1975stochastic}, with time reversed from its terminal-value formulation, gives a unique solution with the regularity of the statement and the representation $u(t,x)=\E_x[b(X_t)\one_{\{\sigma_{D^c}>t\}}]$.
\end{IEEEproof}

\subsection{Proofs of Section~\ref{sec:ltl}}
\begin{IEEEproof}[Proof of Theorem~\ref{thm:monitor}]
(a) We first show, for every finite $\Theta_k$, that $\Lambda_{\Theta_k}$ is consistent with $X_{\Theta_k}$ and with $X_t$ for
$t\in(\Theta_k,\Theta_{k+1})$. By the rules~\eqref{eq:monitor}, the letter at $\Theta_0=0$ is $h\in\Sigma(x)$, and at every finite $\Theta_{k+1}$ it is $\Gamma(\Lambda_{\Theta_k},X_{\Theta_{k+1}})$, which is consistent with $X_{\Theta_{k+1}}$. On $(\Theta_k,\Theta_{k+1})$ the trajectory avoids
$B_{\Lambda_{\Theta_k}}$ by the definition of $\Theta_{k+1}$, and $\Lambda_{\Theta_k}$ is consistent
with $X_t$ there.
(b) By (a), $X_{\Theta_k}\notin B_{\Lambda_{\Theta_k}}$, and since this set is closed and the path is continuous, the next visit happens after a positive time. We fix $t>0$ and a proposition $a$. The image of $[0,t]$ under the trajectory is compact, so if it meets both $A^0_a$ and $A^1_a$, its intersections with them are disjoint compact sets at a positive distance, and by uniform continuity there is $t_a>0$ such that $\lVert X_{t_1}-X_{t_2}\rVert$ is smaller than this distance whenever $\lvert t_1-t_2\rvert<t_a$. Successive switches of $a$ alternate between $A^1_a$ and $A^0_a$ and are therefore at least $t_a$ apart, so $a$ switches at most $1+t/t_a$ times on $[0,t]$, and if the image misses one of the two sets, $a$ switches at most once. Every switching time switches at least one proposition and $\AP$ is finite, so $\lim_k\Theta_k=\infty$, the rules~\eqref{eq:monitor} define $\Lambda$ on $\R_{\ge0}$, and $\Lambda$, being constant on each $[\Theta_k,\Theta_{k+1})$ with finitely many $\Theta_k$ in $[0,t]$, is finitely variable. Thus, (a) holds for every $t\ge0$.
(c) follows from (a) and the two inclusions. The times $\Theta_k$ are hitting times of closed
sets after stopping times, hence stopping times, and $\Lambda$ is constant between them.
\end{IEEEproof}

\begin{IEEEproof}[Proof of Lemma~\ref{lem:band}]
Both sets are closed because the distance is continuous. If $x\in A^1_a$ and $y\in A^0_a$, then
$y\in A_a^c$ and $\lVert x-y\rVert\ge\dist(x,A_a^c)\ge\rho_a$. If $x\in A_a$ and
$\dist(x,\partial A_a)\ge\rho_a$, the segment from $x$ to any $y\in A_a^c$ contains a point of
$\partial A_a$, hence $\lVert x-y\rVert\ge\rho_a$ and $x\in A^1_a$. The case $x\in A_a^c$ is the
same with the roles exchanged.
\end{IEEEproof}

\begin{IEEEproof}[Proof sketch of Lemma~\ref{lem:markov-S}]
The time $\Theta_{k+1}$ is the first visit after $\Theta_k$ to the set $B_{\Lambda_{\Theta_k}}$,
which is determined by $S_k$. The strong Markov property of $X$ at the stopping time $\Theta_k$
gives that the conditional law of $S_{k+1}$ given $\mathcal F_{\Theta_k}$ is $K_\Lambda(S_k,\cdot)$.
\end{IEEEproof}

\begin{IEEEproof}[Proof of Theorem~\ref{thm:ltl}]
By Theorem~\ref{thm:monitor}(b), $\Lambda$ is finitely variable, and its switching times are the
times $\Theta_k$ (Section~\ref{subsec:monitor}). Hence, the $k$-th letter of $\tr(\Lambda)$ is $h_k$
for every finite $\Theta_k$. If the last switch is $\Theta_{k_0}$, then
$h_k=h_{k_0}$ for every $k>k_0$, which is the repetition of the last letter in~\eqref{eq:trace}. The
automaton coordinate is the run of $\mathcal A_\varphi$ on $h_0h_1\cdots$ shifted by one position,
which does not change the priorities visited infinitely often. The process $(S_k,q_k)$ is a DTMP
because $q_{k+1}$ is a deterministic function of $q_k$ and $S_{k+1}$, and
Lemma~\ref{lem:markov-S} completes the proof.
\end{IEEEproof}

\begin{IEEEproof}[Proof of Proposition~\ref{prop:exit-elliptic}]
Fix $h\in\Sigma$ and let $D=B_h^c$. Since $B_h$ is closed and $X$ has continuous paths, a finite
first visit from $D$ occurs on $\partial D$.
(a) Since $X_{\sigma_{B_h}}\in\partial D$ on $\{\sigma_{B_h}<\infty\}$, substituting $\hat b(y)=b(y,\Gamma(h,y))$ in~\eqref{eq:kernel-Lambda} gives both identities.
Choosing $\hat b=1$ in the exit expectation gives the probability of a finite exit.

(b) Let $D_m$ be any bounded open set with $\overline D_m\subset D$. Its exit time satisfies $\sigma_{D_m^c}\le\sigma_{B_h}<\infty$ almost surely, and the strong Markov property gives the mean-value identity $u(x)=\E_x[u(X_{\sigma_{D_m^c}})]$ for $x\in D_m$. By Lemma~\ref{lem:localization}, the coefficients may be replaced on a ball containing $\overline D_m$ by bounded, globally Lipschitz ones without changing the stopped path, so the local regularity argument of~\cite[Thm.~9.2.14]{oksendal2003stochastic} applies to this bounded mean-value function and gives $u\in C^2(D)$ and $\mathcal Lu=0$. The boundary argument of the same theorem, localized near each regular boundary point, gives $u(x)\to\hat b(y)$ as $x\to y$ within $D$, so $u\in C(\overline D)$.

For uniqueness (cf.~\cite[Cor.~9.1.2]{oksendal2003stochastic}), let $\tilde u$ be any bounded classical solution with the prescribed boundary values, and choose smooth bounded sets $D_m$ with $\overline D_m\subset D_{m+1}$ and $\bigcup_mD_m=D$. The stopped generator identity gives $\tilde u(x)=\E_x[\tilde u(X_{t\wedge\sigma_{D_m^c}})]$ for $x\in D_m$, and $\sigma_{D_m^c}\uparrow\sigma_{B_h}$ by nonexplosion and continuity of the paths. Since $\tilde u$ is bounded and continuous on $\overline D$, bounded convergence first as $m\to\infty$ and then as $t\to\infty$, which uses almost-sure exit from every $x\in D$, yields $\tilde u(x)=\E_x[\hat b(X_{\sigma_{B_h}})]=u(x)$.

Finally, let $g g^\top\ge\lambda I$ and let $D$ be bounded with a $C^2$ boundary. For
$c_1>2\sup_{\overline D}\lVert f\rVert/\lambda$, the function $e^{c_1x_1}$ satisfies
$\mathcal Le^{c_1x_1}=e^{c_1x_1}\bigl(c_1f_1+\frac12c_1^2(gg^\top)_{11}\bigr)\ge c>0$ on
$\overline D$, and It\^o's formula up to $t\wedge\sigma_{B_h}$ gives
$c\,\E_x[t\wedge\sigma_{B_h}]\le\sup_{\overline D}e^{c_1x_1}$. Hence, $\E_x[\sigma_{B_h}]<\infty$ and
$\Pr_x(\sigma_{B_h}<\infty)=1$ for every $x\in D$. Moreover, every point of $\partial D$ satisfies an
exterior sphere condition and is therefore regular, by a barrier argument.
\end{IEEEproof}

\subsection{Proofs of Section~\ref{sec:certificates}}
\subsubsection{Proof of Theorem~\ref{thm:reachability}(a)}

\begin{lemma}[Uniform Reach]
\label{lem:uniform-reach}
For every $r\ge0$ and every $s\in\Inv$ with $V(s)\le r$, $\Pr_s\bigl(\sigma_C\le M(r)\bigr)\ge\epsilon(H(r))^{M(r)}$, where $M(r)=\lceil H(r)/d(H(r))\rceil$.
\end{lemma}

\begin{IEEEproof}
Under $\Pr_s$, $Y=(Y_k)_{k\in\N}$ is the chain started at $Y_0=s$. By (R3), $U(s)\le H(r)$. Let $\mathcal E_0=\Omega$ be the sure event, and let
$\mathcal E_{k+1}$ be the event that $\mathcal E_k$ holds and that either $\sigma_C\le k$ or
$Y_{k+1}\in\Inv$ with $U(Y_{k+1})\le U(Y_k)-d(H(r))$. By induction, on $\mathcal E_k$ with
$\sigma_C\ge k$ we have $Y_k\in\Inv$ and $U(Y_k)\le H(r)-k\,d(H(r))$. On $\mathcal E_k$ with
$\sigma_C>k$, the state $Y_k$ lies in $\Inv\setminus C$ with $U(Y_k)\le H(r)$, and (R5) at the level
$H(r)$ gives, conditionally on $Y_0,\dots,Y_k$, a probability at least $\epsilon(H(r))$ for the
decrease. Since $\mathcal E_k$ is determined by $Y_0,\dots,Y_k$ and $\epsilon(H(r))\le1$, this
gives $\Pr_s(\mathcal E_{k+1})\ge\epsilon(H(r))\Pr_s(\mathcal E_k)$.
On $\mathcal E_{M(r)}$, either $\sigma_C<M(r)$, or $U(Y_{M(r)})\le0$ and $Y_{M(r)}\in C$ by (R4).
\end{IEEEproof}

\begin{lemma}[No Escape]
\label{lem:no-escape}
For every $s\in\Inv$, almost surely on $\{\sigma_C=\infty\}$, $Y_k\in\Inv$ for every $k$ and
$\liminf_kV(Y_k)<\infty$.
\end{lemma}

\begin{IEEEproof}
The first claim follows by induction from $K(s',\Inv)=1$ for $s'\in\Inv\setminus C$. For
every $s\in\Inv\setminus C$, the process $V(Y_k)$ stopped at the first visit to $C\cup\Inv^c$ is a
nonnegative supermartingale by (R2), started at the finite value $V(s)$, and therefore converges
almost surely~\cite[Thm.~4.2.12]{durrett2019probability}. On $\{\sigma_C=\infty\}$ the chain never enters
$C$ and, by the first claim, almost surely never leaves $\Inv$. There the stopped process is
$V(Y_k)$ itself, which therefore converges to a finite limit.
\end{IEEEproof}

\begin{IEEEproof}[Proof of Theorem~\ref{thm:reachability}(a)]
By L\'evy's zero-one law~\cite[Thm.~4.6.9]{durrett2019probability}, $\Pr_s(\sigma_C<\infty\mid Y_0,\dots,Y_k)$
converges almost surely to $\one_{\{\sigma_C<\infty\}}$. On
$\{\sigma_C>k,\ Y_k\in\Inv,\ V(Y_k)\le r\}$, the Markov property and Lemma~\ref{lem:uniform-reach}
bound this conditional probability from below by $\epsilon(H(r))^{M(r)}>0$. Hence, almost surely,
the event $\{\sigma_C=\infty\}$ and the event that $Y_k\in\Inv$ and $V(Y_k)\le r$ for infinitely many
$k$ do not occur together. By Lemma~\ref{lem:no-escape}, on $\{\sigma_C=\infty\}$ there is an
integer $r$ with $Y_k\in\Inv$ and $V(Y_k)\le r$ for infinitely many $k$. Taking the union over
$r\in\N$ gives $\Pr_s(\sigma_C=\infty)=0$.
\end{IEEEproof}

\subsubsection{Proof of Theorem~\ref{thm:reachability}(b) and (c)}
\emph{States that reach the target set.} Let $\mathsf S_C$ be the set of states from which $C$ is reached
with probability one. It is Borel, since $\Pr_s(\sigma_C\le k)$ is Borel in $s$ and increases to
$\Pr_s(\sigma_C<\infty)$. For every $s\in\mathsf S_C\setminus C$,
the Markov property gives $1=\int K(s,\dd s')\Pr_{s'}(\sigma_C<\infty)$, with an integrand below one off
$\mathsf S_C$. Hence, $K(s,\mathsf S_C)=1$.

\emph{Layers.} Let $\Xi_0=C$ and $\Xi_j=\Xi_{j-1}\cup\{s\in\mathsf S_C\setminus C:K(s,\Xi_{j-1})>2^{-j}\}$ for $j\ge1$.
These sets are Borel by induction, increase in
$j$ and lie in $\mathsf S_C$. Every state of $\mathsf S_C$ lies in some layer. Otherwise, some $s\in\mathsf S_C\setminus C$ outside all layers has $K(s,\Xi_{j-1})\le2^{-j}$ for every $j\ge1$, hence $K(s,\bigcup_j\Xi_j)=\lim_jK(s,\Xi_j)=0$, and with $K(s,\mathsf S_C)=1$ the next state lies almost surely in $\mathsf S_C$ outside all layers. By induction, the chain started at $s$ never reaches $C$, a contradiction. The variant is $U(s)=\min\{j:s\in\Xi_j\}$ for $s\in\mathsf S_C$ and $U(s)=1$ for $s\notin\mathsf S_C$. It is Borel and vanishes exactly on $C$. A state $s\in\mathsf S_C$ with $U(s)=j\ge1$ lies in
$\Xi_j\setminus\Xi_{j-1}$ and has $K(s,\Xi_{j-1})>2^{-j}$.

\emph{Escape probabilities.} Let $\xi_j(s)=\Pr_s(\sigma_{\Xi_j^c}<\sigma_C)$, Borel in $s$ by
Section~\ref{subsec:model}. Then $\xi_j=1$ on $\Xi_j^c$, $\xi_j=0$ on $C$, $\xi_{j+1}\le\xi_j$, and $K\xi_j\le\xi_j$ on $C^c$.
The first identity holds since, on $\Xi_j^c$, $\sigma_{\Xi_j^c}=0$ and
$\sigma_C\ge1$ by $C\subseteq\Xi_j$, the second since $\sigma_C=0$ on $C$, and the first inequality
since the layers increase. The last is an equality on $\Xi_j\setminus C$ by the
Markov property at time one, and on $\Xi_j^c$ it follows from $\xi_j\le1$. For every $s\in\mathsf S_C$, almost
surely $\sigma_C<\infty$ and $Y_0,\dots,Y_{\sigma_C}\in\mathsf S_C$, by $K(s',\mathsf S_C)=1$ for
$s'\in\mathsf S_C\setminus C$ and induction.
These finitely many states lie in one layer, and $\sigma_{\Xi_j^c}>\sigma_C$ for all large $j$.
Dominated convergence gives $\xi_j(s)\to0$.

\emph{Drift.} Since $\nu$ is finite and $\nu(\mathsf S_C^c)=0$, $\xi_j\to0$ in $\nu$-measure, and there
are indices $j_1<j_2<\cdots$ with $\nu(\{\xi_{j_i}>2^{-i}\})\le2^{-i}$. Let
$V_\infty=\one_{C^c}+\sum_{i\ge1}\xi_{j_i}$, a Borel function with values in $[0,\infty]$, zero on $C$
and at least one on $C^c$. By the Borel--Cantelli lemma~\cite[Thm.~2.3.1]{durrett2019probability}, applied to $\nu/\nu(\mathsf S)$ when $\nu\ne0$, $\nu$-almost every $s$ satisfies
$\xi_{j_i}(s)\le2^{-i}$ for all but finitely many $i$. Since every term of the sum is at most one,
$\nu(\{V_\infty=\infty\})=0$. On $C^c$, monotone convergence, $K(\cdot,C^c)\le1$ and $K\xi_j\le\xi_j$ give
$KV_\infty\le V_\infty$. For $s\notin\Xi_{j_i}$, $\xi_{j_1}(s)=\dots=\xi_{j_i}(s)=1$ and
$V_\infty(s)\ge1+i$. Hence, $V_\infty(s)\le i$ implies $s\in\Xi_{j_i}$ and $U(s)\le j_i$. Also
$V_\infty=\infty$ on $\mathsf S_C^c$, since the layers lie in $\mathsf S_C$.

\emph{Certificate.} Let
\begin{equation*}
  \begin{aligned}
    \Inv&=\{V_\infty<\infty\}, &
    V&=
    \begin{cases}
      V_\infty, & \text{on }\Inv,\\
      1, & \text{on }\Inv^c,
    \end{cases}\\
    H(r)&=j_{\max\{1,\lceil r\rceil\}}, &
    d(r)&=1,\quad \epsilon(r)=2^{-r}.
  \end{aligned}
\end{equation*}
Then $\Inv$ is Borel,
$C\subseteq\Inv\subseteq\mathsf S_C$, $V$ is Borel with $0\le V\le V_\infty$, $H(r)\in\N$ and
$\epsilon(r)\in(0,1]$. For every $s\in\Inv\setminus C$, $(KV_\infty)(s)\le V_\infty(s)<\infty$ gives
$K(s,\Inv)=1$, and $(KV)(s)\le(KV_\infty)(s)\le V(s)$ is (R2). On $\Inv$, $V=V_\infty$ and $U$ vanish
exactly on $C$, which gives (R1) and (R4). For (R3), let $s\in\Inv$ and $r\ge0$ with $V(s)\le r$. Then
$V_\infty(s)\le r\le i$ with $i=\max\{1,\lceil r\rceil\}$, and hence $U(s)\le j_i=H(r)$. For (R5), let
$r\ge0$ and $s\in\Inv\setminus C$ with $j=U(s)\le r$. Then $j\ge1$, every $s'\in\Xi_{j-1}$ has
$U(s')\le j-1=U(s)-d(r)$, and $K(s,\Inv)=1$ gives $K(s,\{s'\in\Inv:U(s')\le j-1\})\ge K(s,\Xi_{j-1})>2^{-j}\ge\epsilon(r)$.
Finally, $\nu(\Inv^c)=\nu(\{V_\infty=\infty\})=0$.

\emph{The second claim of (b) and part (c).} For the second claim of (b), $\Pr_\mu(\sigma_C<\infty)=\int\Pr_s(\sigma_C<\infty)\,\mu(\dd s)=1$ with an integrand at most
one. Hence, $\nu=\mu$ gives no mass to the states from which $C$ is reached with probability less than
one, and the first claim gives a certificate with $\mu(\Inv^c)=0$. For (c), every state reaches $C$ with probability one, and the finite measure
$\nu(A)=\sum_k2^{-k-1}\Pr_\mu(Y_k\in A)$ satisfies the hypothesis of
the first claim of (b). A set of full $\nu$-measure contains $Y_k$ for every $k$,
$\Pr_\mu$-almost surely.

\subsubsection{Proof of Theorem~\ref{thm:streett}(a)}
\begin{lemma}[Hitting at a Positive Time]
\label{lem:return}
Let $I$ and $C$ be Borel sets with $I^c\subseteq C$. Then $\Pr_s(\tau_C<\infty)=1$ for every
$s\in I$ if and only if $\Pr_s(\sigma_C<\infty)=1$, for every $s\in I$.
\end{lemma}

\begin{IEEEproof}
Since $\sigma_C\le\tau_C$, one direction is immediate. For the other, let $s\in I$. The Markov
property at time one gives $\Pr_s(\tau_C<\infty)=\int K(s,\dd s')\Pr_{s'}(\sigma_C<\infty)$.
By assumption, $\Pr_{s'}(\sigma_C<\infty)=1$ for $s'\in I$. For $s'\in I^c\subseteq C$,
$\sigma_C=0$ under $\Pr_{s'}$, and hence $\Pr_{s'}(\sigma_C<\infty)=1$. Therefore,
$\Pr_s(\tau_C<\infty)=\int K(s,\dd s')=1$.
\end{IEEEproof}

\begin{IEEEproof}[Proof of Theorem~\ref{thm:streett}(a)]
By Theorem~\ref{thm:reachability}(a) and (S3), $\Pr_s(\sigma_{C_i}<\infty)=1$ for every $s\in I$.
Hence, by Lemma~\ref{lem:return}, the chain reaches $C_i$ at a positive time from every state of $I$.
By (S2) and~\cite[Thm.~4.1(i)]{abate2026complete}, applied to the complement of $E_i$,
$\Pr_s(\sigma_{E_i}<\infty)\le V_i(s)$ for every $s$. For every $s\in J_i$ this gives
$\Pr_s(\tau_{E_i}<\infty)\le1-\gamma_i$, since $\sigma_{E_i}=\tau_{E_i}$ under $\Pr_s$ for $s\notin E_i$. With
$J_i\subseteq I\setminus E_i$, these are the hypotheses of~\cite[Thm.~3.7]{abate2026complete}, which
bounds $\Pr_\mu(\Str)$ from below by the probability that $Y_k\in I$ for every $k$. By (S1)
and~\cite[Thm.~4.1(i)]{abate2026complete}, the probability of leaving $I$ from $s$ is at most
$V_0(s)$, which gives the bound.
\end{IEEEproof}

\subsubsection{Proof of Theorem~\ref{thm:streett}(b)}
The proof enlarges the target sets of reachability certificates, which the next lemma permits.

\begin{lemma}[Target-Set Enlargement]
\label{lem:goal-enlargement}
Let $(\Inv,V,U,H,d,\epsilon)$ be a reachability certificate for $C$, and let $A\supseteq C$ be a
Borel set. Then $(\Inv\cup A,V\one_{A^c},U\one_{A^c},H,d,\epsilon)$ is a reachability certificate
for $A$.
\end{lemma}

\begin{IEEEproof}
The set $\Inv\cup A$ is Borel and contains $A$. The new functions are Borel, vanish on $A$, are at
most $V$ and $U$, and equal $V$ and $U$ on $A^c$. At a state of $A$ both vanish, and (R1), (R3) and
(R4) for $A$ hold there, since $H\ge0$. Every other state $s$ of $\Inv\cup A$ lies in
$\Inv\setminus C$, where the conditions for $C$ hold. Hence, $K(s,\Inv\cup A)\ge K(s,\Inv)=1$, (R1),
(R3) and (R4) for $A$ hold at $s$, and $(K(V\one_{A^c}))(s)\le(KV)(s)\le V(s)=V(s)\one_{A^c}(s)$. For
$r\ge0$ with $U(s)\le r$, the set $\{s'\in\Inv:U(s')\le U(s)-d(r)\}$ of (R5) for $C$ lies in $\{s'\in\Inv\cup A:U(s')\one_{A^c}(s')\le U(s)\one_{A^c}(s)-d(r)\}$, the set of (R5) for $A$, since $U\one_{A^c}\le U$ and $U(s)\one_{A^c}(s)=U(s)$ for $s\notin A$. Hence, (R5) for $C$ gives (R5) for $A$.
\end{IEEEproof}

Since path probabilities are Borel functions of the initial state (Section~\ref{subsec:model}),
every set and function below is Borel. The results of~\cite{abate2026complete} used below are stated
there for a Markov chain on a general measurable state space~\cite[Sec.~2]{abate2026complete}, and
their first return time is our $\tau$.

\emph{First regions and certificates.} Let $\widehat I$ be the set of states $s$ with
$\Pr_s(\Str)=1$, let $\widehat J_i=\{s\in\widehat I\setminus E_i:\Pr_s(\sigma_{E_i}<\infty)\le1/2\}$,
and let $\widehat C_i=F_i\cup\widehat J_i\cup\widehat I^c$. Since $\sigma_{E_i}=\tau_{E_i}$ under
$\Pr_s$ for $s\notin E_i$, $\widehat J_i$ is the region of~\cite[eq.~(38)]{abate2026complete}. The
event $\Str$ is invariant under time shifts~\cite[Thm.~2.8]{abate2026complete}, and
$\Pr_\mu(\Str)=1$ by hypothesis. Hence~\cite[Lemma~3.13]{abate2026complete}, whose proof uses the
region $\widehat I$, gives $\Pr_\mu(Y_k\in\widehat I\text{ for all }k)=1$. The hypothesis
of~\cite[Lemma~3.9]{abate2026complete}, eq.~(37) there, requires that, for every $s$, $\Pr_s$-almost
surely $\Str$ holds or $Y_k\notin\widehat I$ for some $k$. It holds, since $\Pr_s(\Str)=1$ for
$s\in\widehat I$, and $Y_0=s\notin\widehat I$ otherwise. The proof of that lemma uses the regions
$\widehat J_i$ and shows in~\cite[eq.~(56)]{abate2026complete} that $\Pr_s(\tau_{\widehat C_i}<\infty)=1$ for every
$s\in\widehat I$. Since
$\sigma_{\widehat C_i}\le\tau_{\widehat C_i}$ and $\widehat I^c\subseteq\widehat C_i$, the target set
$\widehat C_i$ is reached with probability one from every state.
Theorem~\ref{thm:reachability}(c) therefore provides a reachability certificate
$\widehat{\mathcal R}_i$ for $\widehat C_i$ whose invariant set $\widehat\Inv_i$ satisfies
$\Pr_\mu(Y_k\in\widehat\Inv_i\text{ for all }k)=1$.

\emph{Final regions.} Let $I$ be the intersection of $\widehat I$ and the sets
$\widehat\Inv_1,\dots,\widehat\Inv_\ell$, let $J_i=\widehat J_i\cap I$, and let $C_i=F_i\cup J_i\cup I^c$. Then $J_i\subseteq I\setminus E_i$, and
$\Pr_\mu(Y_k\in I\text{ for all }k)=1$, as the probability of an intersection of $\ell+1$ events of
probability one. Moreover, $\widehat C_i\subseteq C_i$, since $\widehat I^c\subseteq I^c$ and
$\widehat J_i\setminus I\subseteq I^c$.

\emph{Certificate.} Lemma~\ref{lem:goal-enlargement} with $C=\widehat C_i$ and $A=C_i$ converts
$\widehat{\mathcal R}_i$ into a reachability certificate $\mathcal R_i$ for $C_i$ with the invariant
set $\widehat\Inv_i\cup C_i$, which is $\mathsf S$, since $I\subseteq\widehat\Inv_i$ and
$I^c\subseteq C_i$. This gives (S3). Let $\gamma_i=1/2$, $V_i(s)=\Pr_s(\sigma_{E_i}<\infty)$ and $V_0(s)=\Pr_s(\sigma_{I^c}<\infty)$.
By the Markov property at time one, $(KV_i)(s)=\Pr_s(\tau_{E_i}<\infty)=V_i(s)$ for $s\notin E_i$ and
$(KV_0)(s)=\Pr_s(\tau_{I^c}<\infty)=V_0(s)$ for $s\in I$. Moreover, $V_i=1$ on $E_i$,
$V_i\le1/2$ on $J_i\subseteq\widehat J_i$, and $V_0=1$ on $I^c$. This gives (S1) and (S2), and
$\int V_0\,\dd\mu=\Pr_\mu(\sigma_{I^c}<\infty)=1-\Pr_\mu(Y_k\in I\text{ for all }k)=0$.

\subsubsection{Proof of Theorem~\ref{thm:main}}
Corollary~\ref{cor:emerson-lei} (with $\chi_0=\mathbf 0$ for (b)) with the pairs of Section~\ref{subsec:emerson-lei},
Lemma~\ref{lem:product-Psi}, and~\eqref{eq:ltl-probability} with the pairs $(E_i,F_i)$ of
Section~\ref{sec:certificates} equate the
probability of the requirement with the probability of a Streett event of a DTMP started at a known
state. Theorem~\ref{thm:streett}(a) with the initial distribution concentrated at that state gives
soundness, and Theorem~\ref{thm:streett}(b) gives completeness.

\bibliographystyle{IEEEtran}
\bibliography{SDE}
\begin{IEEEbiography}[{\includegraphics[width=1in,height=1.25in,clip,keepaspectratio]{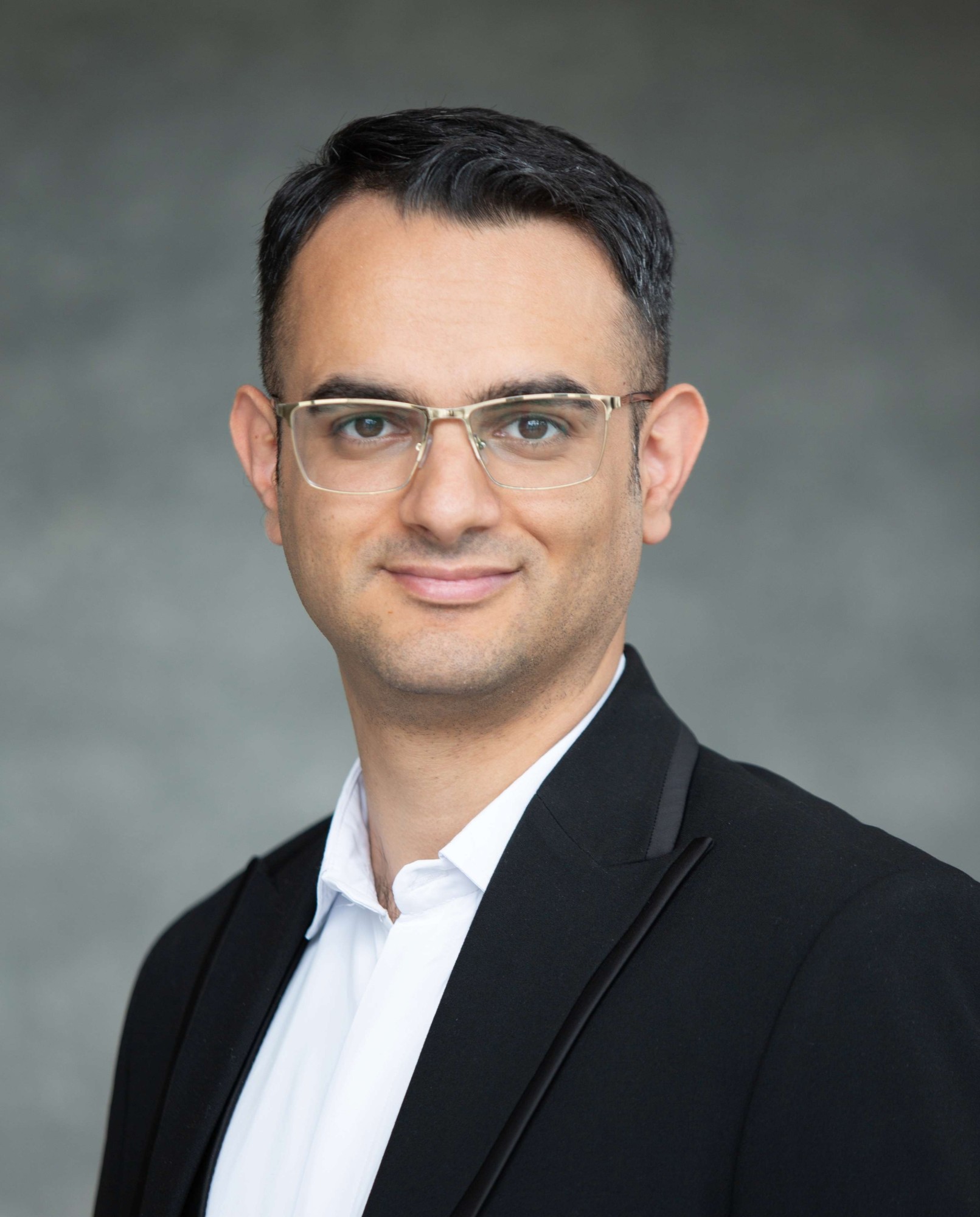}}]{Arash Bahari Kordabad}
received the B.Sc. and M.Sc. degrees in mechanical engineering from the University of Tabriz, Tabriz, Iran, in 2017, and the Sharif University of Technology, Tehran, Iran, in 2019, respectively, and the Ph.D. degree from the Department of Cybernetics Engineering, Norwegian University of Science and Technology (NTNU), Trondheim, Norway, in 2023. He is a Postdoctoral Researcher with the Max Planck Institute for Software Systems, Kaiserslautern, Germany. His research interests include safe reinforcement learning and model predictive control.
\end{IEEEbiography}

\begin{IEEEbiography}[{\includegraphics[width=1in,height=1.25in,clip,keepaspectratio]{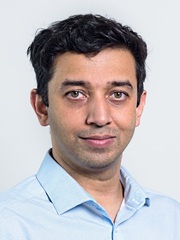}}]{Rupak Majumdar}
received the B.Tech. degree in computer science from the Indian Institute of Technology at Kanpur, Kanpur, India, in 1998, and the Ph.D. degree in computer science from the University of California at Berkeley, Berkeley, CA, USA, in 2003. He is a Scientific Director with the Max Planck Institute for Software Systems, Kaiserslautern, Germany. His research interests are in the verification and control of reactive, real-time, hybrid, and probabilistic systems, software verification and programming languages, logic, and automata theory.
\end{IEEEbiography}

\begin{IEEEbiography}[{\includegraphics[width=1in,height=1.25in,clip,keepaspectratio]{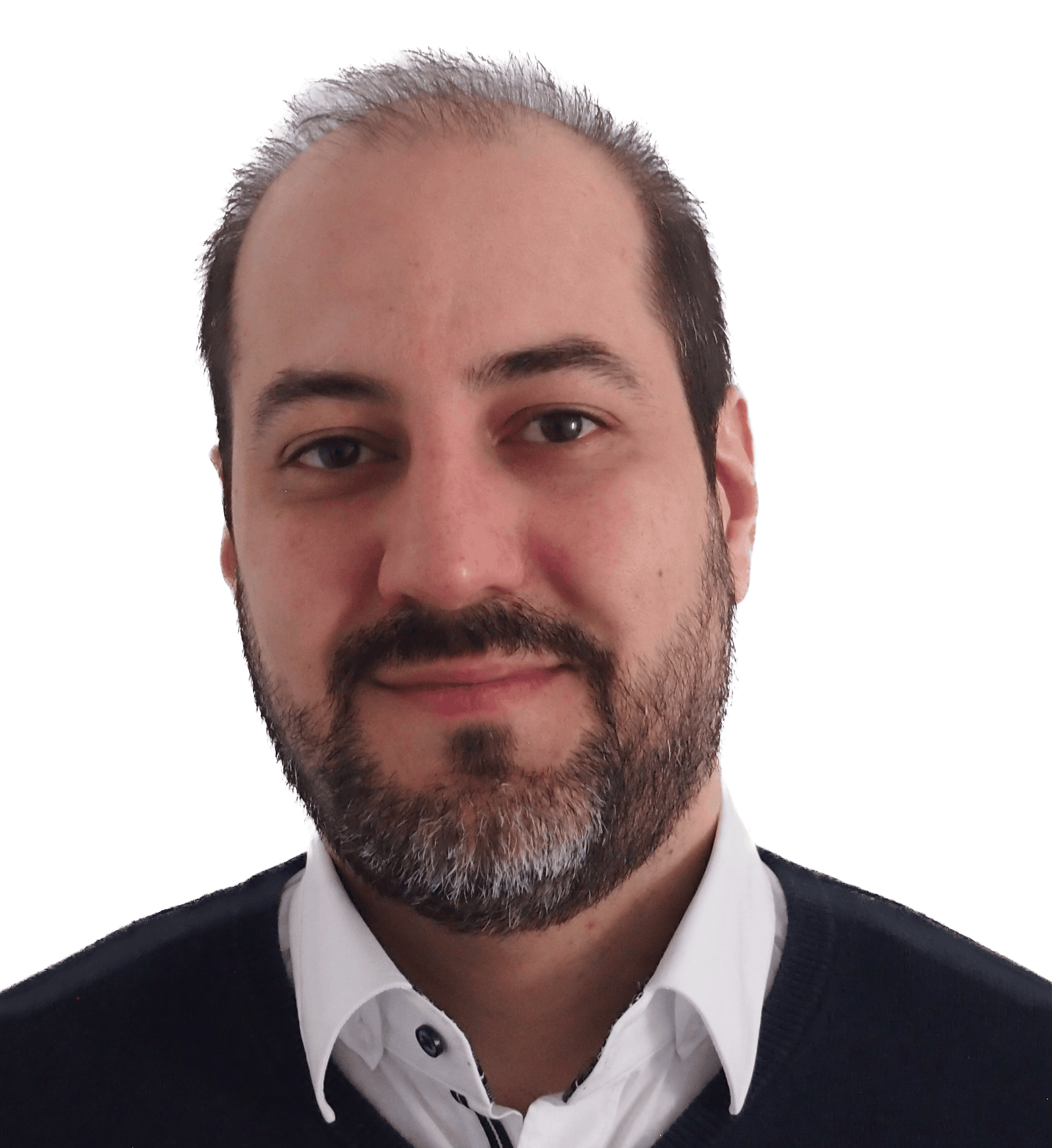}}]{Mohammad H. Mamduhi}
received the Ph.D. degree from the Technical University of Munich (TUM), Germany. He was a Postdoctoral and then a Senior Researcher with KTH Royal Institute of Technology, Sweden, from 2017 to 2020, and a Senior Scientist with ETH Zürich, Switzerland. He is an Assistant Professor with the School of Computer Science, University of Birmingham, U.K., and a Senior IEEE Member. His research interests include networked systems and safe cyber-physical systems.
\end{IEEEbiography}

\begin{IEEEbiography}[{\includegraphics[width=1in,height=1.25in,clip,keepaspectratio]{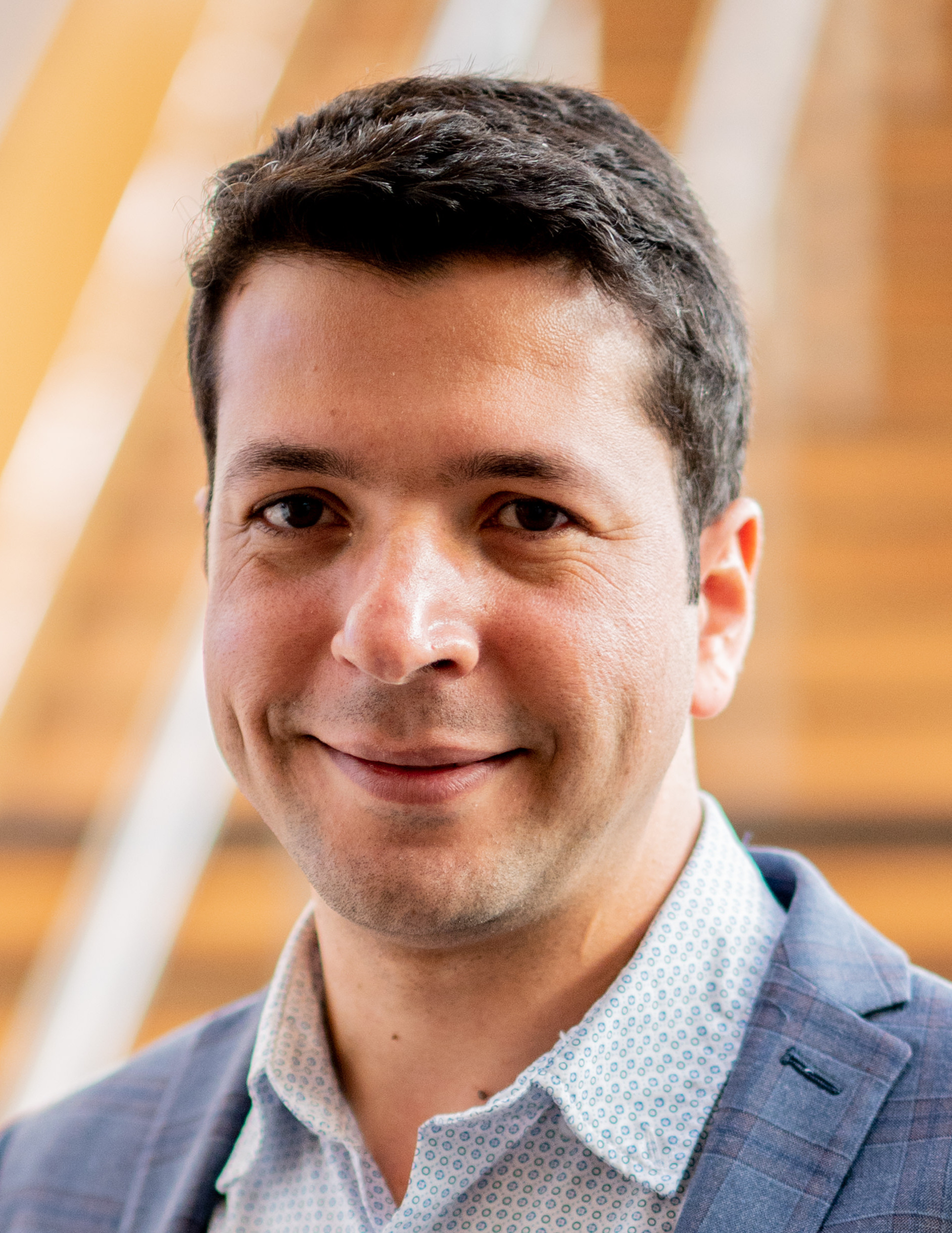}}]{Sadegh Soudjani} received the Ph.D. degree from the Delft Centre for Systems and Control, TU Delft, the Netherlands. He had appointments at Oxford University and Newcastle University. He is Full Professor and Chair in Cyber-Physical Systems at the University of Birmingham, U.K., and a Senior Research Group Leader at the Max Planck Institute for Software Systems, Germany. His research addresses safe autonomy and AI with symbolic and formal techniques for cyber-physical systems. He is the recipient of an ERC Consolidator Grant.
\end{IEEEbiography}
\end{document}